\documentclass[review,12pt]{elsarticle}
\biboptions{numbers,sort&compress}

\usepackage{amssymb}
\usepackage{amsmath}

\usepackage{xcolor}
\usepackage{nomencl}
\makenomenclature

\usepackage{url}  
\usepackage{hyperref} 
\usepackage{enumitem}
\usepackage{booktabs}
\usepackage{array}
\usepackage{tabularx}
\usepackage{ragged2e}
\usepackage{dblfloatfix}
\usepackage{caption}   
\usepackage{graphicx}
\usepackage{tikz}
\newcommand{\hbF}{\tikz[baseline=-0.55ex]{\fill (0,0) circle (0.62ex);}}
\newcommand{\hbH}{\tikz[baseline=-0.55ex]{\draw[line width=0.4pt] (0,0) circle (0.62ex); \fill (0,0) -- (90:0.62ex) arc (90:270:0.62ex) -- cycle;}}
\newcommand{\hbE}{\tikz[baseline=-0.55ex]{\draw[line width=0.4pt] (0,0) circle (0.62ex);}}
\usepackage{float}
\usepackage{pdflscape}
\usepackage{longtable}
\usepackage{placeins}
\usepackage{pdfpages}
\usepackage{atbegshi}


\hypersetup{colorlinks=true, linkcolor=blue, citecolor=blue, urlcolor=blue}

\AtBeginShipout{%
  \global\pdfpagewidth=\paperwidth
  \global\pdfpageheight=\paperheight
}

\newcolumntype{Y}{>{\RaggedRight\arraybackslash}X}

\journal{Energy}

\begin{document}

\begin{frontmatter}

\title{From Net Load Modifiers to Firm Capacity: The Role of Distributed Energy Resources in Resource Adequacy}

\author{Li, Y.}
\author{Moreira, A.}
\author{Heleno, M.\corref{cor1}}
\cortext[cor1]{Corresponding author. Email: MiguelHeleno@lbl.gov}

\affiliation{organization={Lawrence Berkeley National Laboratory},
            addressline={1 Cyclotron Road}, 
            city={Berkeley},
            postcode={94608}, 
            state={CA},
            country={United States}}


\begin{abstract}
Distributed energy resources (DERs) such as rooftop solar, batteries, demand response, and electric vehicles can contribute to power system reliability, yet their performance remains difficult to translate into firm resource adequacy (RA) capacity across jurisdictions. Existing analyses often locate this difficulty within individual technical requirements, such as metering, accreditation, or dispatch performance, but give less attention to how constraints at one stage carry over to the next. This review traces the RA participation pathway through five stages: load forecasting, registration and classification, metering and verification, capacity accreditation, and performance obligations. We synthesize academic literature, tariffs, market manuals, and regulatory documents from California, PJM Interconnection, ISO New England, Great Britain, and Ireland, spanning U.S. capacity markets and European capacity remuneration mechanisms. Across these frameworks, similar participation barriers recur despite different procurement models and regulatory structures, indicating that participation is constrained by cross-stage design rather than by jurisdiction-specific rules alone. We identify three cross-stage couplings through which capacity value is lost between stages: mismatches between resource classification and operational obligations, weak links between verification evidence and accreditation, and temporal misalignment between planning forecasts and scarcity-hour performance. The central finding is that compliance architecture, not DER technology alone, is often the binding constraint on translating distributed-resource capability into firm RA contributions. This points to reforms that codify cross-stage information handoffs, tie accreditation to auditable verification evidence, and refresh capacity values as deployment changes system conditions. Rather than adjusting individual stages in isolation, RA reform should redesign the participation pathway end-to-end.
\end{abstract}

\begin{highlights}
\item A five-stage pathway traces where distributed resources lose capacity credit.
\item Five jurisdictions reveal recurring barriers despite different capacity mechanisms.
\item Entry rules, verification evidence, and scarcity-hour timing shape credited capacity.
\item Cross-stage mismatches, not technology alone, limit how resources count for adequacy.
\item Reforms should coordinate stages rather than fix them in isolation.
\end{highlights}


\end{frontmatter}

\section*{Abbreviations}
\begingroup
\footnotesize
\setlength{\tabcolsep}{4pt}
\renewcommand{\arraystretch}{1.05}
\begin{longtable}{@{}p{2.2cm}p{12cm}@{}}
AMI & Advanced Metering Infrastructure \\
BTM & Behind-the-Meter \\
CAISO & California Independent System Operator \\
CEC & California Energy Commission \\
CHP & Combined Heat and Power \\
CPUC & California Public Utilities Commission \\
CRM & Capacity Remuneration Mechanism \\
DER & Distributed Energy Resource \\
DERP & Distributed Energy Resource Provider \\
DLOL & Direct Loss of Load \\
DNO & Distribution Network Operator \\
DOE & U.S. Department of Energy \\
DR & Demand Response \\
DSO & Distribution System Operator \\
DSU & Demand Side Unit \\
EDC & Electric Distribution Company \\
EE & Energy Efficiency \\
EFC & Equivalent Firm Capacity \\
EFOR & Equivalent Forced Outage Rate \\
ELCC & Effective Load Carrying Capability \\
ERCOT & Electric Reliability Council of Texas \\
EU & European Union \\
EUE & Expected Unserved Energy \\
EV & Electric Vehicle \\
FCM & Forward Capacity Market \\
FERC & Federal Energy Regulatory Commission \\
GDPR & General Data Protection Regulation \\
HVAC & Heating, Ventilation, and Air Conditioning \\
IEPR & Integrated Energy Policy Report \\
IRP & Integrated Resource Planning \\
ISO & Independent System Operator \\
ISO-NE & ISO New England \\
LDA & Locational Deliverability Area \\
LIP & Load Impact Protocol \\
LOLE & Loss of Load Expectation \\
LOLH & Loss-of-Load Hours \\
LSE & Load-Serving Entity \\
LTRA & Long-Term Reliability Assessment \\
MISO & Midcontinent Independent System Operator \\
NEM & Net Energy Metering \\
NERC & North American Electric Reliability Corporation \\
NYISO & New York Independent System Operator \\
PDR & Proxy Demand Resource \\
PDR-LSR & Proxy Demand Resource-Load Shift Resource \\
PJM & PJM Interconnection \\
PRD & Price Responsive Demand \\
PRM & Planning Reserve Margin \\
PV & Photovoltaic \\
RA & Resource Adequacy \\
RDRR & Reliability Demand Response Resource \\
RTO & Regional Transmission Organization \\
SCADA & Supervisory Control and Data Acquisition \\
Sub-LAP & Sub-Load Aggregation Point \\
TSO & Transmission System Operator \\
UCAP & Unforced Capacity \\
UDC & Utility Distribution Company \\
VPP & Virtual Power Plant \\
VRE & Variable Renewable Energy \\
\end{longtable}
\addtocounter{table}{-1}
\endgroup




\section{Introduction}

\subsection{Motivation and Practical Relevance}

RA is formally defined as the ability of the electric system to supply aggregate electrical demand and energy requirements at all times, taking into account scheduled and reasonably expected unscheduled outages of system elements \cite{doe2025resource}. Historically, this objective was pursued in a paradigm dominated by large, centralized, dispatchable generators and relatively passive demand. Adequacy assessment therefore evolved around ensuring sufficient bulk-level firm generation capacity to meet aggregate demand under stochastic generator and transmission outages, typically within vertically integrated utility structures or explicit CRMs. 

However, this bulk-system-centered foundation is now structurally challenged by concurrent shifts on both the supply and demand sides of the adequacy equation. 
On the supply side, accelerated retirements of aging thermal plants are shrinking the pool of conventional firm capacity, while increasing VRE penetration introduces correlated availability constraints that shift reliability risk across more hours and seasons \cite{nerc_ltra_2024}. On the demand side, widespread electrification of transportation and heating is projected to increase peak demand and alter load shapes, often stressing local distribution infrastructure \cite{doe2024future}.
Simultaneously, the rapid expansion of energy-intensive data centers is increasing both baseload demand and the value of lost load, adding pressure on existing adequacy standards \cite{shehabi2024united}. These trends are further compounded by more frequent and severe extreme weather events \cite{nerc_ltra_2024}, which invalidate historical planning assumptions and increase the probability of hours where available capacity margins fall below reliability thresholds.

Against this backdrop, DERs are reshaping both sides of the adequacy equation. 
DERs comprise a heterogeneous class of distribution-connected and BTM resources, including rooftop PV, battery storage, demand response, controllable loads, and grid-interactive EVs, whose collective capacity is sufficient, in principle, to contribute materially to system adequacy \cite{Forrester2021}. 
When coordinated through aggregation platforms or virtual power plant architectures, DERs can provide dispatchable load reduction and injection capacity, reducing net system peaks and deferring bulk infrastructure investment \cite{muhtadi2021distributed}. 
While regulatory reforms such as FERC Order No. 2222 have established formal wholesale market access for DER aggregations \cite{FERC_Order_2222_2020}, translating their technical flexibility into firm, accredited capacity remains structurally constrained.

The underlying mismatch is structural: existing RA frameworks were designed for centralized generators with deterministic, transmission-observable availability, yet DERs present fundamentally different characteristics across the entire adequacy lifecycle, from net-load forecasting and interconnection, through accreditation, to market procurement and operational enforcement. 
In net-load forecasting, millions of behind-the-meter devices are deployed by decentralized customer decisions rather than utility planning, introducing structural uncertainty that is difficult to bound probabilistically \cite{maheshwari2020effect}. 
In accreditation, performance is heterogeneous and state-dependent, from battery state-of-charge to demand-response capacity inferred from counterfactual baselines subject to contamination bias \cite{9701332}. 
In deliverability, feeder-level thermal and voltage limits can block accredited capacity during loss-of-load events even when transmission-level models indicate full deliverability, systematically overstating firm capacity \cite{oikonomou2019deliverable}. 
Finally, registration rules, telemetry standards, and performance obligations calibrated for centralized generators force conservative de-rating or outright exclusion of DER portfolios \cite{ESIG_DER_Integration_2022}. 


\subsection{Limits of Existing Literature}

Existing literature has illuminated pieces of the DER adequacy problem but rarely treats the translation pathway as a connected institutional process. Planning and accreditation studies have advanced probabilistic methods such as ELCC for variable generation, storage, and demand response \cite{billinton1970power,garip2022power,nerc2012methods,stenclik2023ensuring,carvallo2023guide}, but often assume bulk-system deliverability and abstract from registration, telemetry, baseline, and enforcement rules \cite{carvallo2023guide,leibowicz2024importance,ESIG_DER_Integration_2022}. Control and optimization work demonstrates the feasibility of DER aggregation and VPP operation \cite{muhtadi2021distributed,ochoa2023control}, yet commonly assumes frictionless market integration rather than the compliance constraints that filter flexibility before capacity-market clearing \cite{ESIG_DER_Integration_2022,Huang2025_VPP_ResponseCapability}. Market-design and regulatory studies document CRM outcomes or DER access rules \cite{botterud2020resource,Lynch2019_DR_CapacityMarkets,EPRI_DERA_OrganizedMarkets_2021,EldridgeSomani2022_FERC2222}, but less often ask whether sequential rules successfully convert physical capability into credited and enforceable capacity.

These strands leave a specific gap: they explain parts of the DER adequacy problem, but not how forecasting, entry, metering, accreditation, and enforcement interact as a pathway. Table~\ref{tab:review_scope} shows that recent reviews tend to emphasize one or two functions, such as probabilistic assessment, CRM design, market access, local flexibility, or VPP operations, while leaving the full planning-to-enforcement chain weakly connected. This review addresses that gap by covering all five dimensions, comparing five jurisdictions, and analyzing how constraints at one stage propagate into binding limits at another.

\begin{table}[htbp]
\linespread{1}\selectfont
\caption[Scope comparison with recent reviews]{Scope of recent reviews compared with this review. \protect\hbF~substantial, \protect\hbH~partial, \protect\hbE~not addressed. Functional dimensions: Planning, Entry, Metering, Accreditation, Enforcement. Cross-cutting dimensions: Jurisdictions and Interactions.}
\label{tab:review_scope}
\footnotesize
\setlength{\tabcolsep}{3pt}
\renewcommand{\arraystretch}{1.1}
\begin{tabularx}{\textwidth}{@{} >{\RaggedRight\arraybackslash}p{2.2cm} >{\RaggedRight\arraybackslash}X *{7}{>{\centering\arraybackslash}p{0.75cm}} @{}}
\toprule
& & \multicolumn{5}{c}{\textbf{\shortstack{Functional\\dimensions}}} & \multicolumn{2}{c}{\textbf{\shortstack{Cross-\\cutting}}} \\
\cmidrule(lr){3-7}\cmidrule(lr){8-9}
\textbf{Review} & \textbf{Primary focus}
 & \rotatebox{90}{Planning}
 & \rotatebox{90}{Entry}
 & \rotatebox{90}{Metering}
 & \rotatebox{90}{Accreditation}
 & \rotatebox{90}{Enforcement}
 & \rotatebox{90}{Jurisdictions}
 & \rotatebox{90}{Interactions} \\
\midrule
Botterud \& Auer, 2020 \cite{botterud2020resource} & EU vs.\ U.S.\ CRM design under rising VRE & \hbH & \hbH & \hbE & \hbH & \hbH & \hbF & \hbE \\
Gulotta et al., 2023 \cite{gulotta2023opening} & Opening ancillary-service markets to DERs & \hbE & \hbF & \hbH & \hbE & \hbH & \hbF & \hbE \\
Stenclik et al., 2023 \cite{stenclik2023ensuring} & Capacity-accreditation design for high-renewable systems & \hbH & \hbE & \hbH & \hbF & \hbH & \hbH & \hbE \\
Carvallo et al., 2023 \cite{carvallo2023guide} & Guide to RA assessment in evolving systems & \hbF & \hbH & \hbE & \hbF & \hbH & \hbH & \hbE \\
Leibowicz et al., 2024 \cite{leibowicz2024importance} & Operational detail in probabilistic RA & \hbF & \hbE & \hbE & \hbF & \hbE & \hbE & \hbH \\
Gao et al., 2024 \cite{Gao2024_VPP_APEN} & VPP operations and resource coordination & \hbE & \hbH & \hbH & \hbE & \hbE & \hbE & \hbE \\
Haugen et al., 2024 \cite{Haugen2024_MarketModels_APEN} & Power-market models for the clean-energy transition & \hbF & \hbE & \hbE & \hbH & \hbH & \hbH & \hbE \\
Menci \& Valarezo, 2024 \cite{Menci2024_LFM_APEN} & Taxonomy of local flexibility markets & \hbE & \hbF & \hbH & \hbH & \hbH & \hbF & \hbE \\
Mantegna et al., 2024 \cite{mantegna2024comprehensive} & Capacity-accreditation methods across resource types & \hbH & \hbE & \hbE & \hbF & \hbE & \hbH & \hbE \\
Sanchez Jimenez et al., 2025 \cite{SanchezJimenez2025_CRMs_APEN} & Agent-based evaluation of CRMs & \hbH & \hbH & \hbE & \hbH & \hbH & \hbH & \hbE \\
Tillmanns et al., 2026 \cite{Tillmanns2026_ProbabilisticRA} & Probabilistic RA assessment methods & \hbF & \hbE & \hbE & \hbF & \hbE & \hbH & \hbE \\
\midrule
\textbf{This review} & \textbf{End-to-end pathway for distributed resources to count toward adequacy} & \hbF & \hbF & \hbF & \hbF & \hbF & \hbF & \hbF \\
\bottomrule
\end{tabularx}
\end{table}

\subsection{Objective and Contributions}

This review examines how RA frameworks determine whether DER capability can be counted toward adequacy. To study this systematically, we organize the evidence around a five-stage pathway: an upstream planning layer followed by entry and classification, metering and verification, capacity accreditation, and obligations and enforcement. The review synthesizes recent academic literature with implementation evidence from technical reports, regulatory filings, tariffs, and business practice manuals. The paper makes three contributions:

\begin{itemize}[leftmargin=*,itemsep=2pt,topsep=2pt]
\item First, we introduce the five-stage value-translation pathway as an analytical framework for tracing how DER capability becomes accredited capacity. Rather than treating DER barriers as isolated technical problems, the framework identifies where capability is preserved, discounted, or excluded as it passes through forecasting, entry, verification, accreditation, and enforcement.
\item Second, we conduct a document-grounded comparison of California, PJM, ISO-NE, Great Britain, and Ireland to assess whether DER participation barriers persist across different RA procurement and enforcement designs. The recurrence of similar barriers across these cases indicates that the constraints are structural features of RA design rather than artifacts of a single jurisdiction.
\item Third, we identify three cross-stage couplings that explain why reforms at individual stages often fail to scale DER participation: entry categories shape downstream obligations, verification evidence constrains accredited capacity, and forecast timing affects whether credited capacity matches scarcity-hour performance. These couplings lead to three design principles: make information handoffs explicit, base accreditation on auditable verification evidence, and update capacity credits as deployment and performance conditions change.
\end{itemize}

\paragraph{Review scope and sources} The evidence base combines academic literature with regulatory and market documentation, including tariffs, business practice manuals, compliance filings, and technical reports. Jurisdictions are selected for having explicit RA or capacity-remuneration frameworks, codified DER participation rules, and sufficient public documentation, and for spanning centralized and decentralized designs: California, PJM, ISO-NE, Great Britain, and Ireland.

The remainder of the paper is organized as follows. Section 2 clarifies the RA architecture and design features most relevant to DER participation. Section 3 introduces the analytical framework used in the review, combining a DER role taxonomy with a five-stage pathway. Section 4 applies that framework in a gate-by-gate review of current practices, barriers, and design implications. Section 5 synthesizes the cross-stage coupling mechanisms that limit scalable participation and derives coordination principles for reform. Section 6 concludes the paper.

\begin{figure}[!htbp]
  \centering
  \makebox[\textwidth][c]{\includegraphics[width=1.1\textwidth]{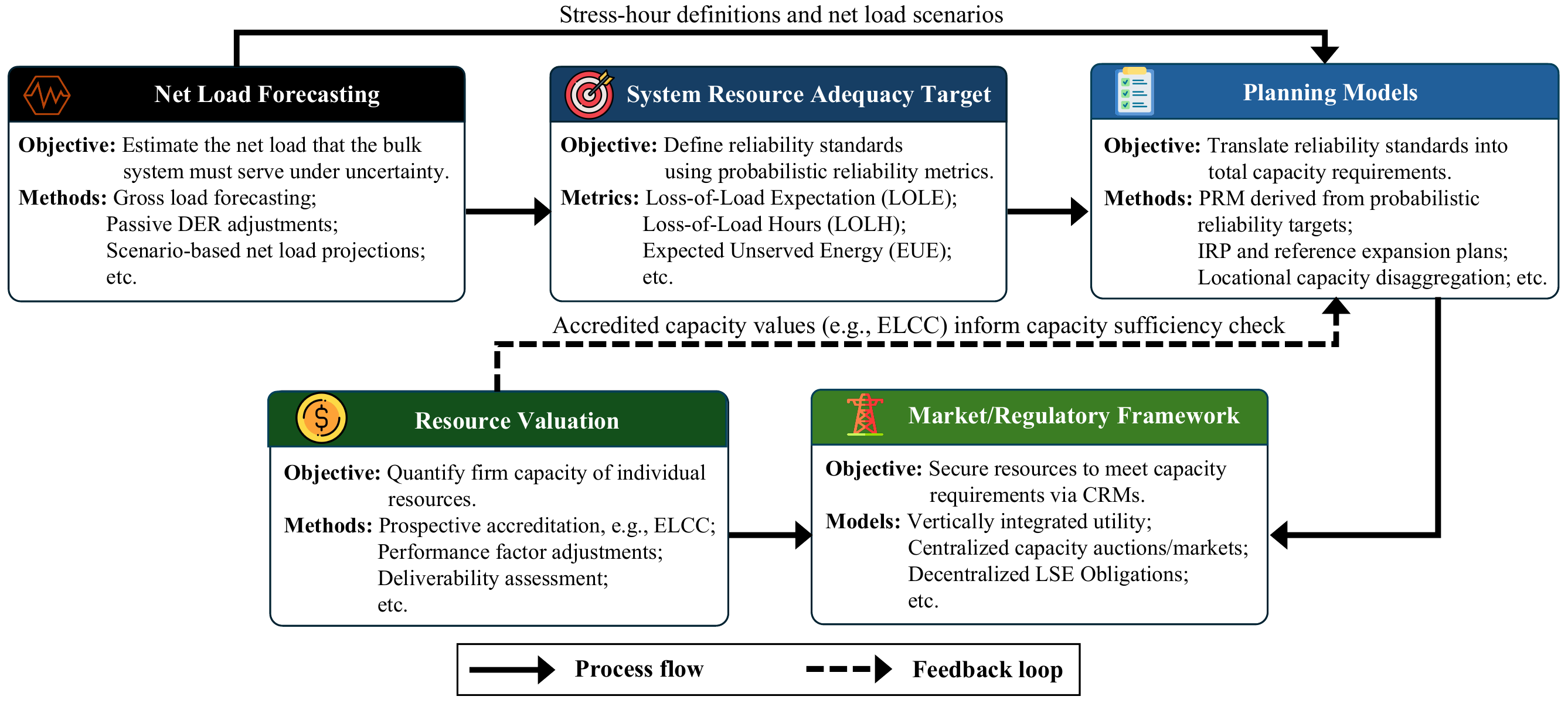}}
  \caption{Functional architecture of conventional resource adequacy frameworks, linking net-load assumptions, reliability targets, capacity accreditation, and procurement obligations. The figure distinguishes the upstream planning logic from the downstream institutional logic through which DER participation is later evaluated.}
  \label{fig:ra_framework}
\end{figure}

\section{Resource Adequacy Architecture and Institutional Design}
\label{sec:traditional_ra_frameworks}

\subsection{Core Principles of RA Assessment}
Understanding why DER participation does not scale in RA requires first clarifying how existing frameworks convert reliability targets into measurable, accredited, and enforceable capacity obligations. Figure~\ref{fig:ra_framework} summarizes the conventional planning-to-compliance chain: planners estimate the net load to be served, define an acceptable loss-of-load risk, translate that risk into a capacity requirement, accredit individual resources against that requirement, and secure commitments through centralized auctions or decentralized procurement obligations \cite{nerc_glossary_2016}. The early stages define the adequacy problem, while the later stages determine what capacity can actually be counted and enforced. In restructured markets, explicit CRMs make this chain visible because eligibility, accreditation, and enforcement rules are codified rather than internal to a utility planning process \cite{botterud2020resource}. This review therefore focuses on CRM-based or binding RA frameworks, where the institutional filtering of DER capacity value can be traced directly.

\subsection{Reliability Targets and Capacity Requirements}

Reliability targets define the adequacy need that later accreditation and procurement must satisfy. Legacy planning often summarized this need through a deterministic planning reserve margin, but fixed peak-based margins are less informative when weather-dependent and energy-limited resources create correlated outages, duration limits, and multi-period scarcity conditions \cite{billinton1970power,nerc2012methods,5374085,Tillmanns2026_ProbabilisticRA}. Modern RA studies therefore increasingly use probabilistic metrics: LOLE remains a common target metric, while LOLH describes expected scarcity hours and EUE measures the expected magnitude of unserved energy \cite{nerc2012proba,nerc_ltra_2024,carvallo2023guide,Tillmanns2026_ProbabilisticRA}.

Many frameworks still report the result as a planning reserve margin, but planners increasingly choose a reliability target first and then estimate the reserve margin needed to meet it. Requirements may also be disaggregated by location, season, or ramping need; California's separate system, local, and flexible RA obligations illustrate this structure \cite{caiso2025bpm,CPUC_RA_Landing}. This step matters for DERs because passive DERs reshape the net-load forecasts used to set the target, while active DER accreditation is calibrated against the resulting stress hours, locations, and operating conditions.

\subsection{Resource Capacity Accreditation}
\label{sec:accreditation}

System-level studies determine how much total capacity the system requires; accreditation determines how many megawatts a specific resource can count toward that requirement. It is therefore the operational link between probabilistic reliability targets and resource-specific procurement decisions.



A conventional thermal generator is commonly accredited as UCAP, de-rating nameplate capacity by EFOR to reflect outage availability \cite{nerc2012methods}. This logic fits dispatchable units with roughly independent failures, but not VRE, storage, or demand response. VRE output is weather-correlated, storage depends on duration and state of charge, and demand response depends on baselines and repeated-call behavior. Many jurisdictions have therefore moved toward ELCC-style accreditation using stochastic, chronological simulations to estimate each resource class's marginal reliability contribution \cite{pjm2021elcc,nyiso2025icap,MISO_DLOL_2024}.

For energy-limited resources, accreditation must distinguish operating constraints, so 2-hour, 4-hour, and 8-hour storage devices can receive different capacity values \cite{nerc_ltra_2024,MISO2024,wang2021crediting,gonzato2023effect}. For demand-side resources, accredited capacity often depends on baseline methods, load-impact studies, and performance or availability adjustments that reflect call limits, fatigue, and rebound \cite{CPUC_LIP_2024,Lynch2019_DR_CapacityMarkets,stenclik2023ensuring}. These methods are more resource-sensitive than static de-rates, but their outputs still depend on upstream net-load scenarios, stress-hour definitions, deliverability assumptions, and baseline methodologies \cite{carvallo2023guide,stenclik2023ensuring}.

\subsection{Market Frameworks and Regulatory Context for RA}

Reliability targets and accreditation methods become consequential only through institutions that procure capacity and enforce obligations. Since electricity restructuring, RA in many jurisdictions has moved from internal utility planning to explicit CRMs or binding procurement requirements. Centralized capacity markets, including PJM, ISO-NE, NYISO, and MISO, assign procurement to the system operator, which runs forward auctions to meet an aggregate reliability requirement and attaches performance obligations to cleared resources \cite{PJM_M18,ISO_NE_M20_2023,nyiso2025icap,MISO2024}. Decentralized obligation frameworks, such as California, assign requirements to LSEs, which procure bilaterally or self-supply while planning agencies and system operators verify need, deliverability, and backstop procurement \cite{CPUC_RA_Landing,caiso_lca_study_2026,cec2025iepr,CAISO_TPD}.

The institutional form differs, but the planning-to-compliance sequence is similar: resources must qualify under an administrative category, provide evidence that can be verified, receive an accredited capacity value, and accept obligations that can be settled or penalized. This sequence was designed around large, centrally coordinated, dispatchable resources whose performance could be observed through transmission-level metering and treated separately across planning, valuation, and operations \cite{nerc2012methods}. These assumptions enabled a clean separation between reliability assessment, capacity valuation, and compliance, but they create challenges for DERs whose capability must pass through the same sequence before it can count as RA capacity.

\section{Analytical Framework: Translating DER Capability into Accredited Capacity} 
The assumptions identified in Section~2 map directly onto the DER problem. Centralized planning gives way to customer-driven adoption and aggregator coordination; transmission-level observability gives way to baseline-inferred or state-dependent performance; and sequential planning, accreditation, and enforcement are complicated by distribution-level deliverability constraints. Rather than treating these as isolated technical barriers, this review examines the pathway through which DER capability is represented, credited, or filtered out before becoming accredited capacity.

This section defines that framework. Section~3.1 distinguishes passive load modifiers from active capacity resources and groups active DERs by how performance is measured and credited. Section~3.2 then maps the pathway from planning assumptions to enforceable obligations. The purpose is to identify where DER capability is preserved, discounted, or excluded before it can satisfy RA requirements.

\subsection{DER Roles in Resource Adequacy}

DERs play two roles in RA. Passive DERs alter the net-load target without registering as capacity resources or accepting delivery obligations; active DERs are enrolled, accredited, and committed as resources whose performance can be verified and enforced. The same device can appear in either role depending on contracts, telemetry, and regulatory authorizations. Because netting DERs into load can obscure distinct risk profiles, we distinguish resources by how RA rules recognize their contribution rather than by ownership or voltage level \cite{nerc2012methods,carvallo2023guide}.

\paragraph{Passive DERs: Shaping the net load baseline}
Passive DERs include unregistered BTM PV, small CHP, energy efficiency, and other behind-the-meter measures represented as load or net-load adjustments rather than separately accredited capacity resources \cite{nerc2012methods,carvallo2023guide,CPUC_LIP_2024,cec2025iepr}. 
Passive DERs shape the stress-hour definitions and procurement targets against which active resources are later evaluated.

\paragraph{Active DERs: Enrolled capacity resources with RA obligations}
Active DERs are resources that enter RA with an accredited capacity value and a delivery obligation. For cross-jurisdictional comparison, we classify them by how their performance is measured and credited, because registration as generation, demand response, storage, or aggregation determines the verification, accreditation, and enforcement rules applied downstream \cite{PJM_M18,caiso2020bpmdr,ISO_NE_M20_2023}.

First, \textit{supply-side DERs} are distribution-connected resources that inject power and are registered and metered as capacity, including small generators, community solar, distribution-connected batteries, and BTM generation or storage aggregations. They are usually treated as generator-like or storage-like assets subject to deliverability tests and telemetry requirements. For storage, the accredited contribution also depends on state of charge and duration rather than time-invariant availability \cite{caiso2025bpm,pjm2021elcc,MISO_DLOL_2024,wang2021crediting,gonzato2023effect}.

Second, \textit{load-modifying capacity resources} reduce or shift demand relative to an estimated baseline during scarcity conditions, including curtailable loads, managed EV charging, and other flexible demand. Because avoided consumption cannot be observed directly, accreditation depends on baseline and load-impact methods, with adjustments for performance, repeated-call limits, and rebound \cite{CPUC_LIP_2024,Lynch2019_DR_CapacityMarkets,9701332,stenclik2023ensuring}. The same flexibility can therefore receive different RA value under different baseline windows and event definitions.

Third, \textit{hybrid portfolios} combine supply-side and load-modifying capabilities within one operational portfolio, such as PV-plus-storage plants or VPPs coordinating BTM PV, batteries, thermostats, and managed EV charging \cite{Huang2025_VPP_ResponseCapability,Zhu2025_HybridEMS,Dykes2020_HybridPlants,DOE_2023_VPP_Liftoff,Liu2024_VPP_DynamicAggregation}. Many frameworks still require these portfolios to register under legacy generation, storage, or demand-response categories \cite{FERC_Order_2222_2020,EPRI_DERA_OrganizedMarkets_2021,Ericson2022_HybridELCC}. This decomposition can misrepresent coordinated response capability and tie accreditation to administrative boundaries rather than portfolio-level physical behavior \cite{PJM_M18,caiso2020bpmdr,Ericson2022_HybridELCC,stenclik2023ensuring}. Hybrid portfolios therefore make the review's central argument most visible: what limits credited value is institutional classification, not the portfolio's physical capability.

\begin{figure}[htbp]
\centering
\makebox[\textwidth][c]{\includegraphics[width=1.3\textwidth]{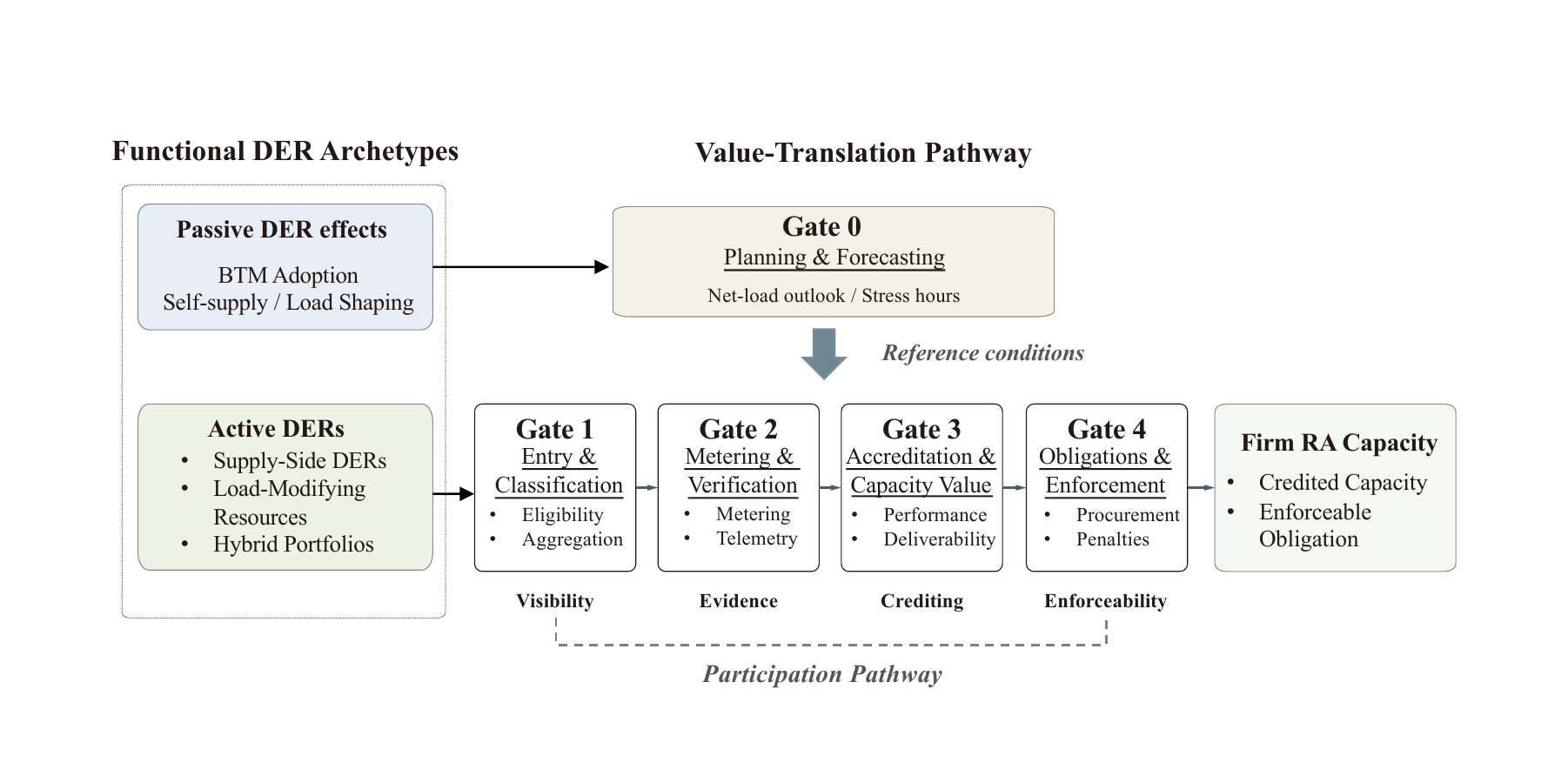}}
\caption{DER roles and the five-stage value-translation pathway. Passive DERs shape net-load forecasts and stress-hour definitions, while active DERs pass through entry, verification, accreditation, and enforcement before their capability becomes credited and enforceable RA capacity.}
\label{fig:der_framework}
\end{figure}

\subsection{Five Stages from Net-Load Effects to Accredited Capacity}

Active DERs must pass through a structured process before physical capability becomes accredited capacity. Although RA designs differ across jurisdictions, the progression from planning assumptions to enforceable obligations is broadly similar. We use five stages to locate where DER capability is preserved, discounted, or lost.

\begin{enumerate}[leftmargin=*, label=\textbf{G\arabic*}, start=0]
    \item \textbf{Planning and Forecasting:} Passive DER adoption shapes demand forecasts, net-load scenarios, risk-hour definitions, and reliability targets. These outputs set the reference conditions against which active DER capacity is later accredited and enforced.
    \item \textbf{Entry and Classification:} Interconnection and market-qualification rules determine whether DER capability enters RA as a recognizable compliance unit. The assigned participation category determines which verification, accreditation, and obligation rules apply downstream.
    \item \textbf{Metering and Verification:} Telemetry, metering, baseline, and data-governance rules define the evidence that can be treated as auditable capacity. Measurement therefore varies by participation model: supply-side resources generally rely on direct injection or net-output metering, load-modifying resources require baseline-based verification, and hybrid portfolios usually combine direct metering with calculated or component-level evidence rather than a single uniform protocol.
    \item \textbf{Accreditation and Capacity Value:} Accreditation converts verified or assumed capability into capacity credit, subject to deliverability and availability constraints. Supply-side resources are typically credited through reliability-based methods such as ELCC, load-modifying resources through load-impact estimates and performance adjustments, and hybrid portfolios often through component-level aggregation.
    \item \textbf{Obligations and Enforcement:} Accredited capacity becomes a binding commitment defined by product windows, penalty rules, and availability requirements. For DER portfolios, enforcement also depends on how overlapping wholesale, retail, and distribution obligations are prioritized and how network-directed curtailment is attributed.
\end{enumerate}

Resources that fail at any stage may still reduce net load, but they do not count as accredited capacity toward the RA requirement.


\section{Gate-by-Gate Review of DER Participation in Resource Adequacy}
\label{sec:gate_by_gate}
This section applies the five-stage framework to current DER participation practice. Table~\ref{tab:challenges_directions} provides the roadmap by summarizing the main challenges and design directions, while the subsections supply the supporting evidence. Gate~0 draws on adequacy-planning studies and related academic literature; Gates~1--4 draw on implementation documents and prior research, with Supplementary Table~S1 serving as a compact summary of the five-jurisdiction comparison.

\begin{table}[htbp]
\caption{Key challenges and design directions across the five-stage pathway.}
\label{tab:challenges_directions}
\fontsize{8}{8.5}\selectfont
\setlength{\tabcolsep}{3pt}
\renewcommand{\arraystretch}{1.0}
\begin{tabularx}{\textwidth}{@{} p{2.4cm} Y Y @{}}
\toprule
\textbf{Stage} & \textbf{Key challenges} & \textbf{Design directions} \\
\midrule
\textbf{G0}\newline
\textit{Planning and forecasting}
& \begin{itemize}[leftmargin=*,nosep]
    \item Forecasts can lag DER deployment, shifting stress hours before delivery.
  \end{itemize}
& \begin{itemize}[leftmargin=*,nosep]
    \item Version DER assumptions, stress-hour definitions, and update triggers.
    \item Record DER trajectories alongside net-load profiles.
  \end{itemize} \\
\midrule
\textbf{G1}\newline
\textit{Entry and classification}
& \begin{itemize}[leftmargin=*,nosep]
    \item Size, location, and approval rules can fragment portfolios.
    \item Legacy categories can lock hybrids into mismatched downstream rules.
  \end{itemize}
& \begin{itemize}[leftmargin=*,nosep]
    \item Use shared qualification records across utilities, regulators, and system operators.
    \item Let mixed aggregations qualify as single compliance units.
  \end{itemize} \\
\midrule
\textbf{G2}\newline
\textit{Metering and verification}
& \begin{itemize}[leftmargin=*,nosep]
    \item Baseline verification makes load reductions counterfactual and bias-prone.
    \item Telemetry costs and data-access limits constrain auditability.
  \end{itemize}
& \begin{itemize}[leftmargin=*,nosep]
    \item Report baseline error and measurement uncertainty.
    \item Use role-based, privacy-preserving audit data.
  \end{itemize} \\
\midrule
\textbf{G3}\newline
\textit{Accreditation and capacity value}
& \begin{itemize}[leftmargin=*,nosep]
    \item Transmission deliverability can miss distribution constraints.
    \item Static DR and component de-rates can miss fatigue, rebound, clustering, and portfolio interactions.
  \end{itemize}
& \begin{itemize}[leftmargin=*,nosep]
    \item Feed distribution deliverability limits into RA models.
    \item Use state-dependent and portfolio-level accreditation.
  \end{itemize} \\
\midrule
\textbf{G4}\newline
\textit{Obligations and enforcement}
& \begin{itemize}[leftmargin=*,nosep]
    \item Stacked obligations can overcommit the same assets.
    \item Settlement can confuse non-delivery with network curtailment.
  \end{itemize}
& \begin{itemize}[leftmargin=*,nosep]
    \item Track commitments with priority rules.
    \item Log curtailment and reconcile shortfalls after events.
  \end{itemize} \\
\bottomrule
\end{tabularx}
\end{table}

\subsection{Gate 0: Net-load formation and planning representation}
\label{subsec:gate0}

Gate~0 concerns how DERs are represented in the adequacy problem before market participation begins. It is an upstream planning layer rather than a participation screen: it sets the reference conditions under which downstream gates evaluate whether active DER capability can be counted in RA.

\subsubsection{Passive DER representation in adequacy planning}

In jurisdictions with explicit RA constructs, passive DER effects are typically incorporated through exogenous adjustments to baseline demand. BTM solar, energy efficiency, EV charging, and distributed storage are commonly represented as scenario-based load modifiers that reduce gross demand and generate net-load inputs for adequacy assessments \cite{BestPractices_IRP_2024}. The resulting hourly profiles are then used to identify reliability-critical hours and to inform capacity procurement targets.


Planning documents illustrate this logic in different settings: California's IEPR foregrounds duck-curve and evening-ramp stress, NERC's LTRA stress-tests adequacy under extreme weather and electrification, and European outlooks embed country-specific DER trajectories \cite{cec2025iepr,cpuc_irp_2024,nerc_ltra_2024,entsoe_mat_2023}. These exercises typically compress DER uncertainty into a reference forecast and a small number of alternatives, which then shape the stress hours used in procurement and accreditation. As DER penetration rises, adequacy risk is increasingly defined by shifting net-load stress periods rather than legacy gross-peak conditions alone.


\subsubsection{Key challenges}
The recurring challenge at Gate~0 is the lag between forecast vintages and realized DER deployment. The lag is visible in the administrative calendar: California RA load-forecast materials are submitted roughly a year before the compliance year, while NERC's 2024 LTRA applies demand and resource data submitted and validated in July 2024 to a 2025--2034 assessment window \cite{CPUC_RA_Landing,nerc_ltra_2024}. These forecast vintages are necessary for procurement and modeling, but DER adoption after the forecast date can change the stress hours used to size obligations and accredit resources \cite{cec2025iepr,BestPractices_IRP_2024}. In high-DER systems, continued BTM solar growth can shift scarcity from afternoon gross-load peaks toward evening net-load ramps \cite{cole2020considerations,cec2025iepr}, so capacity value can become misaligned before any explicit de-rating rule is applied. Because Gate~0 fixes the reference conditions, a mis-specified net-load or stress-hour definition propagates into every later gate.

\subsubsection{Design directions}

Gate~0 design should treat each forecast vintage as a versioned reference package. That package should document the DER adoption assumptions used in each scenario, the mapping from gross load and passive DERs to net-load stress hours, and the conditions under which a material divergence from observed DER deployment triggers an update. Where DER penetration is high, RA planning should keep the distributed-generation and storage trajectories used to build each net-load profile as part of the forecast record, consistent with guidance that these resources and demand carry different uncertainty profiles even when they are netted together for RA assessment \cite{BestPractices_IRP_2024}. This gives downstream accreditation and procurement a common reference case when stress hours shift between planning and delivery.

\subsection{Gate 1: Entry and Classification}
\label{subsec:gate1}

Gate~1 determines whether heterogeneous DER capability can enter the RA framework as a recognizable compliance unit, and under which participation model it will be evaluated. The issue is therefore not access alone, but whether a DER portfolio is admitted in a form that remains measurable, eligible for accreditation, and enforceable downstream.

\subsubsection{Qualification pathways and cross-jurisdictional practice}

Qualification rules perform three linked functions. They set the minimum size and technical evidence required for entry, define the geographic footprint over which assets may be aggregated, and assign the registered portfolio to a participation category such as demand response, storage, generation, or mixed DER. FERC Order~2222 requires ISOs and RTOs to establish DER aggregation participation models with minimum-size requirements no larger than 100~kW, locational rules, metering and telemetry rules, and coordination among the ISO/RTO, aggregator, distribution utility, and relevant retail regulator \cite{FERC_Order_2222_2020,FERC_Order_2222A_2021}; EU Directive 2019/944 establishes independent aggregator access rights while leaving DSO coordination to Member States \cite{EU_Directive_2019_944,JRC_ExplicitDR_2022}. These access rules broaden formal eligibility, but the assigned participation category still determines the accreditation, obligation, and settlement rules applied later.

Current implementations show why formal access does not by itself preserve portfolio-level capability. California PDR, RDRR, and PDR-LSR registrations are built around CAISO Sub-LAPs, with LSE and UDC review before CAISO approval; PDR and RDRR performance must exclude measured export, and only the PDR-LSR curtailment resource ID is eligible for RA \cite{caiso2020bpmdr}. PJM's DER capacity registration rules require locations to remain within the same LDA, transmission zone, state, and EDC territory, and a location cannot support overlapping DER capacity, load management, PRD, peak-shaving, or non-retail BTM generation registrations for the same period \cite{PJM_M18}. ISO-NE defines an Active Demand Capacity Resource as demand-response resources in the same Dispatch Zone, registered with ISO-NE for FCM obligations \cite{ISO_NE_Order2222_Conforming_2025}. European frameworks set higher entry floors and different boundaries: Great Britain's Capacity Market sets a 1~MW minimum within a DNO license area, with non-firm connections exposed to curtailment, while Ireland's Demand Side Units enter at 4~MW under dual TSO and market-operator approval \cite{UK_CM_Flex_Response_2025,EirGrid_DSU_2026}. Thus, a battery, flexible load, or BTM generator that can be controlled by the same aggregator may still be excluded from the same RA registration if it sits in a different qualifying area, belongs to an incompatible retail program, or provides a response that the selected participation model does not recognize.

\subsubsection{Key challenges}

\paragraph{Size, location, and approval rules can fragment portfolios before qualification}

The first entry barrier is procedural rather than technological. A controllable portfolio can only be registered to the extent that its component locations fit the boundary and approval tests of the chosen participation model. In California, PDR, RDRR, and PDR-LSR registrations must remain within a CAISO Sub-LAP, and the LSE and UDC review the location information before CAISO approval \cite{caiso2020bpmdr}. In PJM, DER capacity registrations are constrained by LDA, transmission zone, state, and EDC territory, with multi-nodal registration allowed only within those limits and only where the locations cannot separately meet the 0.1~MW threshold \cite{PJM_M18}. In ISO-NE, Active Demand Capacity Resources must be composed of demand-response resources in the same Dispatch Zone \cite{ISO_NE_Order2222_Conforming_2025}.

These rules determine the MWs that are countable for RA. An aggregator may control a larger set of devices, but the RA process recognizes only the locations that pass the relevant boundary, registration, and capacity-qualification checks. PJM codifies this through nominated MWs from locations registered and approved in DR Hub before the delivery year, while California requires PDR/RDRR/PDR-LSR curtailment resources to complete registration and establish NQC before they enter RA or supply plans \cite{PJM_M18,caiso2020bpmdr}. The binding constraint is therefore not technical flexibility but legibility: whether a dispersed portfolio can be rendered visible to market software and system operators as a single compliance unit \cite{Huang2025_VPP_ResponseCapability,Zhu2025_HybridEMS}.

\paragraph{Participation category fixes the downstream measurement and crediting rules}

The second challenge is category fit. Once a portfolio is registered, the participation model determines which performance evidence can be used for capacity qualification and compliance. In California, PDR and RDRR are load-curtailment products: performance is measured against an approved baseline, measured export from individual locations is excluded, and only the PDR-LSR curtailment resource ID is eligible for RA \cite{caiso2020bpmdr}. In PJM, the same physical location cannot support overlapping DER capacity, load management, Price Responsive Demand, peak-shaving, or non-retail BTM generation registrations for the same period \cite{PJM_M18}. In ISO-NE, an Active Demand Capacity Resource is composed of Demand Response Resources within the same Dispatch Zone and is registered for FCM obligations \cite{ISO_NE_Order2222_Conforming_2025}. These rules create different measurement and crediting pathways, not different names for the same resource.

The implication for hybrid and VPP portfolios is concrete. If one portfolio combines load reduction, charging flexibility, discharge, and BTM generation, only the functions recognized by the registered participation model enter the RA calculation. Load-curtailment pathways rely on baselines and event performance, whereas supply-side or storage pathways rely on interval metering, injection or withdrawal records, availability, and duration \cite{caiso2020bpmdr,PJM_M18,ISO_NE_Order2222_Conforming_2025}. Without a portfolio-level model, coordinated response is either split across component registrations or reduced to the part compatible with one category \cite{Dykes2020_HybridPlants,DOE_2023_VPP_Liftoff,Huang2025_VPP_ResponseCapability}. Gate~1 can therefore cap credited RA value before performance is verified or accredited, because the accepted evidence has already been narrowed by the selected participation model.

\subsubsection{Design directions}

These challenges indicate that reforms focused solely on lowering formal entry barriers are insufficient. Shared digital qualification infrastructure (unified asset registries with standardized telemetry and verification evidence packages reusable across interconnection, retail enrollment, and RA registration) can reduce sequential approval friction and multi-party coordination costs \cite{Horowitz2019_DERInterconnection,Brown2023_ESI_Requirements,IEA2022_UnlockingDER}. Scalable participation ultimately requires participation models that recognize heterogeneous portfolios as primary compliance units, with qualification tests and obligation structures that reflect integrated response capability \cite{DOE_2023_VPP_Liftoff,Dykes2020_HybridPlants}. The appropriate criterion for evaluating Gate~1 design is not whether formal eligibility is broad, but whether the qualification pathway preserves coordinated capability for downstream measurement, accreditation, and enforcement.

\subsection{Gate 2: Metering and Verification}
\label{subsec:gate2}

Gate~2 determines what evidence can verify DER capability and therefore what later gates can treat as auditable capacity; for distributed resources this is not only a metering problem but one of evidence quality and data governance.

\subsubsection{Verification pathways and current observability models}

Current practice combines multiple observability pathways rather than requiring a single uniform verification standard for all DERs. This approach typically operationalizes observability through three distinct mechanisms: (i) telemetry exemptions for capacity-only resources, where operators waive real-time SCADA requirements for smaller aggregations in favor of settlement-quality interval metering to reduce entry barriers \cite{caiso2020bpmdr,ISO_NE_M20_2023}; (ii) ex-post verification regimes that assess compliance retrospectively against scarcity events rather than through real-time monitoring \cite{PJM_M18}; and (iii) inferential visibility for load-modifying resources, where capacity contributions are estimated via counterfactual baselines or static peak contributions rather than direct physical measurement \cite{PJM_M18,CPUC_LIP_2024}. Observability requirements tighten with size and product scope: CAISO requires telemetry for Proxy Demand Resources at or above 10~MW or providing reserves \cite{caiso2020bpmdr}, while aggregation-oriented pathways rely on aggregate metering, as in Great Britain's Satisfactory Performance Days tested at half-hourly aggregation rather than device-level telemetry \cite{UK_CM_Flex_Response_2025,Ofgem_CM_AnnualReport_2024_25,UK_CM_Scheme_Update_2025}; per-jurisdiction detail appears in Table~S1. Taken together, these arrangements suggest a practical tradeoff between broader participation and granular observability for resources not providing ancillary services.

Across jurisdictions, these mechanisms add up to a consistent pattern: the more verification is inferred, aggregated, or event-triggered rather than directly and continuously observed, and the less it is auditable across aggregators, distribution utilities, and system operators, the more conservatively downstream accreditation must treat the result.


\subsubsection{Key challenges}

\paragraph{DER adequacy contributions are often inferential rather than directly observable}
A first recurring challenge is that DER capacity contributions are often inferential rather than directly observable. For load-modifying resources in particular, delivered capacity must be estimated relative to a counterfactual baseline rather than read directly from a meter, making verification dependent on baseline construction, event selection, and behavioral assumptions that cannot be validated prospectively \cite{PJM_M18,CPUC_LIP_2024}. Baseline methods can introduce directional bias in both directions: randomized control trial evaluations show that behavioral spillover can cause standard methods to underestimate the actual peak load reduction delivered by 39--46\% \cite{todd2019spillover}, while anticipatory pre-event load shifting can inflate estimated baselines in the opposite direction \cite{CPUC_LIP_2024}. More broadly, a review spanning eight baseline methodology families and twenty-three European field deployments finds that no single approach provides universally satisfactory accuracy, and that baseline design imposes a persistent trade-off among computational complexity, implementation cost, and estimation bias \cite{Valentini2022_CBL_Review}.

\paragraph{Verification architectures trade observability against scalable participation}
A second, distinct challenge arises even where performance can be directly observed: stronger observability frequently comes at the cost of scalable participation. Device-level metering, generator-style telemetry, and high-frequency data requirements can improve deterministic visibility, but they also impose per-site instrumentation and communications costs that scale poorly for mass-market DER portfolios \cite{DOE_2023_VPP_Liftoff,EldridgeSomani2022_FERC2222}. This tension is visible in current market design: FERC Order~2222 proceedings documented broad stakeholder concern that applying revenue-grade telemetry standards to individual small-scale DERs within aggregations would be economically prohibitive, motivating exemptions such as CAISO's sub-10~MW telemetry waiver noted above \cite{caiso2020bpmdr,EPRI_DERA_OrganizedMarkets_2021}. Frameworks therefore rely on telemetry exemptions, aggregation-level metering, and ex-post testing pathways to preserve economic viability, yet these same accommodations reduce evidentiary quality and lead to more conservative downstream treatment. Verification architecture thus determines which portfolios receive capacity credit: Great Britain's Capacity Market applies a uniform 79\% de-rating to all demand-side resources regardless of demonstrated response quality, while MISO's resource accreditation reform assigns class-level default values where individual performance evidence is insufficient \cite{UK_CM_Flex_Response_2025,MISO_DLOL_2024}.

\paragraph{Verification is increasingly an inter-organizational data architecture problem}
A third challenge is that verification is increasingly an inter-organizational data architecture problem rather than a single metering problem. Adequacy-relevant evidence is distributed across aggregators, DSOs, and system operators, while privacy regulations such as GDPR and CCPA can limit ISO access to granular behind-the-meter data needed to validate baselines or audit compliance \cite{EU_Directive_2019_944,Kua2023_AMI_PrivacySurvey,CPUC_DER_Action_Plan_2022,EU_Implementing_Reg_2023_1162}. This creates a black-box verification problem: performance may be technically plausible yet institutionally difficult to validate end-to-end, and current trust-but-verify audits lack scalability for high-penetration futures \cite{Psara2022_EU_DDS,JorgensenMa2025_DT_EU,EU_Data_Act_2023}. In practice, Gate~2 sets the evidentiary ceiling for Gate~3 and the auditability ceiling for Gate~4: limited observability constrains the quality of downstream accreditation, while fragmented auditability weakens later settlement, attribution, and non-performance accountability.

\subsubsection{Design directions}

These challenges suggest that Gate~2 reform should focus less on extending generator-style telemetry to all DERs and more on verification architectures that make uncertainty explicit and auditable. Recent wholesale-market aggregation shows that scalable participation is achievable through portfolio-level verification with controlled availability risk: aggregators derive supply curves with disclosed uncertainty bounds from offline sampling of actual usage data, reported to meet ISO compliance standards while preserving customer privacy in a demonstration \cite{Papalexopoulos2024_VPP}, and feeder-level deep-learning methods have estimated aggregate DER capacity to about 97.5\% accuracy in a test system without device-level telemetry \cite{Nikzad2024_AggregateDER}. Operationalizing these principles requires standardized data-sharing procedures with role-based access: aggregators retain customer data, DSOs provide delivery evidence, and system operators receive aggregated performance attestations \cite{Brown2023_ESI_Requirements,EU_Data_Act_2023}. The design criterion is not whether every device can be made fully visible, but whether distributed performance can be documented with evidence credible enough for downstream accreditation and enforcement.

\subsection{Gate 3: Accreditation and Capacity Value}
\label{subsec:gate3}
Gate~3 converts observed or assumed DER capability into accredited capacity that can count toward RA requirements. It is not a neutral accounting step: the accredited value depends on benchmark choice (marginal vs.\ average, stress-hour weighting), chronology treatment, portfolio interaction assumptions, and deliverability screening, not solely on the resource's physical characteristics. The same physical capability can therefore be over-credited, aligned, or under-credited depending on the method applied.

\subsubsection{Current practices}
Gate 3 accreditation largely inherits the technology-based registration and obligation logic established in Gate 1 entry and classification. Once resources enter the RA framework through class-specific rules, accreditation typically converts heterogeneous physical capability into standardized capacity credits using qualification tests and performance adjustments.

\paragraph{Supply-side accreditation through reliability metrics and deliverability screening}
For supply-side resources in U.S. capacity markets, accreditation is increasingly tied to reliability contribution metrics such as ELCC, implemented through technology-specific de-rating curves, seasonal factors, and test-based mappings from nameplate to qualified capacity \cite{caiso2025bpm, pjm2021elcc,MISO_DLOL_2024,PJM_M18}. Deliverability is assessed in parallel through ISO and RTO transmission power flow studies \cite{PJM_M14B_2024,CAISO_Deliv_Issue}, while distribution constraints are handled through separate utility interconnection processes rather than through RA deliverability assessments directly.

\paragraph{Demand-side accreditation through baseline methodologies}

For load-modifying resources, accreditation remains anchored in baseline and performance methodologies, with qualifying capacity inferred from measured load reductions relative to counterfactual baselines \cite{nyiso2025icap,CPUC_LIP_2024,ISONE_OP14_2025,MISO_BPM011_2025}. Great Britain's Capacity Market illustrates the class-level character of this approach: all demand-side resources receive a single 79\% de-rating factor regardless of response duration or portfolio composition \cite{UK_CM_Flex_Response_2025,Ofgem_CM_AnnualReport_2024_25,UK_CM_Scheme_Update_2025}. Where ELCC-style methods are applied to demand-side resources, they more often appear as class-level calibration layered on baseline-based nominations rather than as full replacements for measurement-based accreditation \cite{PJM_M18,CPUC_IRP_Incremental_ELCC_2023}.

\paragraph{Portfolio accreditation through component aggregation}

For mixed portfolios, participation is increasingly organized through VPP-like aggregations \cite{DOE_2023_VPP_Liftoff}, but most RA frameworks still form portfolio credits by summing class-level accredited values rather than accrediting the portfolio as a primary unit \cite{PJM_M18,cpuc2025sod}. Even where aggregation-level capability testing exists, as in NYISO's DER Aggregation model, underlying technology accreditation remains class-based without probabilistic or ELCC-style adjustments \cite{NYISO_DER_Aggregation_FAQ}. Chronological reliability simulations in IRP settings increasingly evaluate composite portfolios with portfolio-level ELCC representations \cite{CPUC_IRP_Incremental_ELCC_2023}, though these planning outputs have not yet translated into portfolio-first accreditation rules for RA compliance.

Accreditation choices across the five jurisdictions are compared in Table~S1 in the supplementary material.

\subsubsection{Key challenges}

\paragraph{Transmission-focused deliverability screening can overstate accredited capacity}

A first recurring challenge is that accreditation frameworks assign capacity value based primarily on transmission deliverability assessments while omitting distribution-level constraints that may bind during system stress \cite{PJM_M14B_2024,CAISO_Deliv_Issue}. Feeder and substation limits are handled through separate interconnection and hosting-capacity processes rarely integrated into RA deliverability frameworks \cite{HCAReview2023}. Distribution studies show that bulk-only approximations can materially overestimate export capability: time-series analyses incorporating voltage and thermal limits show that feeder-level export capability can fall substantially below nameplate ratings on constrained feeders, with deliverable headroom varying widely across feeders with similar DER penetration \cite{taheri2020fast,cuenca2021non}.

This separation creates two forms of over-accreditation. First, capacity may be accredited for nameplate or test-based ratings yet remain undeliverable during risk hours due to feeder congestion, creating a gap between accredited and physically deliverable capacity \cite{Frick2021_LocationalValue}. Second, static deliverability caps applied uniformly across all hours fail to capture state-dependent coincidence between local congestion and system scarcity. Distribution limits that bind primarily during high solar output or moderate load may not constrain capacity during the extreme evening or multi-day stress conditions that drive LOLE, yet uniform de-rates treat all hours equally \cite{taheri2020fast}. This creates a risk of over-crediting capacity stranded behind distribution bottlenecks during the conditions that most threaten adequacy.

\paragraph{Demand-side accreditation remains highly sensitive to baseline construction and chronology-dependent availability}

A second challenge is that demand-side capacity value is often accredited using static or historically averaged adjustments even though actual availability is chronology-dependent and behaviorally state-dependent \cite{carvallo2023guide,nethercutt2023resource}. Baseline-driven accreditation raises two main problems. First, baseline contamination: methods developed for infrequent, independent events infer counterfactuals from recent non-event consumption, but as dispatch frequency increases, prior events, pre-event load shifting, and tariff-driven behavioral changes contaminate recent history \cite{todd2019spillover,smith2025potential}. Randomized control trial evaluations show that behavioral spillover can cause standard baseline methods to underestimate actual peak load reduction delivered by 39--46\%, while anticipatory pre-event load shifting can inflate estimated baselines in the opposite direction \cite{todd2019spillover,CPUC_LIP_2024}. California LIP evaluations document this failure mode under frequent calls and dynamic pricing \cite{CPUC_LIP_2024}.

Second, availability degrades systematically under the multi-day stress conditions that drive adequacy risk: demand response availability falls across consecutive stress days due to participation fatigue and post-event rebound \cite{nerc_ltra_2024,Sun2023_CoolingPatterns_Snapback,NREL_DR_Fatigue_Snapback_2024}, while managed EV charging weakens when events cluster due to mobility constraints \cite{CEC2019_SmartChargingEV}. Yet frameworks typically accredit these resources using static capacity values averaged over historical single-event performance, rarely propagating fatigue, rebound, or multi-day clustering effects into LOLE and EUE calculations \cite{carvallo2023guide,nethercutt2023resource}. Omitting these effects can significantly overestimate system reliability, producing over-accreditation concentrated precisely during the multi-day stress periods that most threaten adequacy \cite{mantegna2024comprehensive}.


\paragraph{Component-based accreditation can misrepresent coordinated portfolio value and create double counting}
A third challenge is that mixed DER portfolios are still commonly accredited through component-based rules that cannot capture coordinated portfolio behavior or overlapping contributions \cite{PJM_M18,cpuc2025sod}. This creates two distinct distortions. First, summing independent technology-level de-rates misses correlated availability and joint operating constraints: when storage, demand response, and flexible loads serve the same customers or feeders, a battery may be depleted precisely when demand response is also fatigued after consecutive dispatch events, so the portfolio delivers less than the sum of individually accredited components \cite{qi2023portfolio,sun2021research}.

Second, internal double counting arises when multiple components receive separate credits for overlapping net-load effects: PV-driven reductions depress demand-response baselines, yet both PV and DR are accredited separately, and storage shifting energy across hours creates portfolio-level effects not uniquely attributable to individual components \cite{carvallo2021implications,yuan2025flexibility}.

Component-based accreditation therefore creates a gap between compliance metrics and actual deliverable capacity, concentrating shortfall risk when correlated under-delivery materializes across dependent resources.


\subsubsection{Design directions}


\paragraph{Feeder-aware deliverability for distribution-connected resources}

To address distribution-level deliverability gaps, Gate 3 accreditation would need to internalize feeder and substation constraints through auditable data exchanges between distribution planning and RA modeling \cite{HCAReview2023}. This does not require embedding full feeder physics in LOLE studies. Instead, the aim is to translate local constraints into reduced-form deliverability limits, such as feeder-specific de-rates, stress-hour adjustment factors, or zone-level capacity caps, that are stable, updateable through routine distribution planning cycles, and consumable by reliability models \cite{Keen2022_DistCapacityExpansion}. 

Implementation can leverage existing distribution analysis without full co-simulation: probabilistic hosting-capacity studies already quantify export limits under load, output, and contingency uncertainty \cite{HCAReview2023,taheri2020fast}; locational capacity-value frameworks for non-wires alternatives translate network constraints into node- or zone-specific adjustment factors \cite{Frick2021_LocationalValue}.

\paragraph{State-dependent availability models for demand-side accreditation}

For baseline construction, evidence favors segment-specific baseline distributions informed by AMI and device telemetry over single deterministic counterfactuals, with published data-quality flags and aggregation-level error statistics making uncertainty auditable without exposing customer-level traces \cite{Bakare2023_DSM_AMI,Sun2019_ResidentialBaseline}. For reliability modeling, chronological simulations would treat demand response as a stochastic process whose state-dependent availability, driven by weather, event frequency, and dispatch history, propagates persistence decay and rebound directly into LOLE and EUE calculations rather than applying static de-rates \cite{Lynch2019_DR_CapacityMarkets}.

\paragraph{Portfolio-level accreditation}
To address joint operating constraints, correlated availability, and double counting, mixed portfolios would be better accredited as primary units where institutional conditions permit. Chronological reliability simulations with Monte Carlo sampling track joint state trajectories in which energy limits, fatigue, and rebound evolve across consecutive stress hours, rather than summing independent technology de-rates, capturing the correlated behavior that component-level methods miss \cite{esig2021,leibowicz2024importance,Wang2024_MultiTimeScale_CapCredit,WenSong2023_WindStorage_ELCC}. Emerging VPP evidence demonstrates calibration of portfolio response distributions using aggregated AMI and device telemetry, enabling auditable updates to persistence assumptions without exposing individual customer data \cite{DOE_2023_VPP_Liftoff}. 

For valuation, portfolio-level marginal ELCC provides a single capacity value reflecting net portfolio effects, avoiding ambiguous attribution when components contribute jointly to overlapping net-load reductions \cite{Ericson2022_HybridELCC}. More generally, the relevant question at Gate~3 is not how to assign a single static capacity number to each DER, but how to keep accredited capacity aligned with the portfolio's actual reliability contribution under the stressed multi-day conditions that most threaten adequacy. Accredited capacity values from Gate~3 become the basis for binding obligations and enforcement at the final gate.

\subsection{Gate 4: Obligations and Enforcement}
\label{subsec:gate4}

Gate~4 translates accredited capacity into binding commitments and determines how performance shortfalls are assessed during system stress. It therefore decides whether upstream capacity value is enforceable in practice, especially when portfolios face overlapping obligations or distribution-driven curtailment.

\subsubsection{Current practices}
\paragraph{Technology-neutral obligation structures}

Many jurisdictions implement broadly technology-neutral enforcement frameworks. U.S. organized markets such as PJM and ISO-NE use annual capacity products with pay-for-performance settlement during scarcity intervals, while NYISO structures obligations around summer and winter capability periods \cite{nyiso2025icap, PJM_M18,ISO_NE_PFP_2025}. California's Slice-of-Day framework applies finer temporal granularity, translating adequacy needs into hourly showing requirements better aligned with storage and load-modifying characteristics \cite{cpuc2025sod}. 
In ISO-NE, pay-for-performance settlement is calculated in five-minute Capacity Scarcity Condition intervals \cite{ISO_NE_PFP_2025}. ISO-NE's 2024 Annual Markets Report records 65 minutes of total 10-minute reserve scarcity and 130 minutes of total 30-minute reserve scarcity during the year, illustrating how relatively short scarcity intervals can nevertheless create material performance exposure \cite{ISONE_AMR_2024}. European implementations also settle against system-stress events: Great Britain's Capacity Market penalizes capacity agreements that underdeliver during stress events, while Ireland's reliability options require capacity providers to refund market prices above a strike price during scarcity \cite{UK_CM_Scheme_Update_2025,Ofgem_CM_AnnualReport_2024_25,SEM_CRM_Performance_2022}.

However, DER portfolios may participate across multiple programs simultaneously and may also face distribution-level operating constraints during bulk-system scarcity hours, creating enforcement ambiguities that existing frameworks address through ad-hoc coordination rather than systematic design.

\subsubsection{Key challenges}

\paragraph{Multi-program participation without unified accounting}

DER portfolios provide value across multiple markets simultaneously: wholesale capacity, retail demand response, ancillary services, and distribution-level non-wires alternatives \cite{Frick2021_LocationalValue,ShenotEtAl2024_PVPlusDERs,gulotta2023opening}. However, existing frameworks lack mechanisms to track and reconcile overlapping commitments. Each program maintains separate enrollment, dispatch protocols, and performance assessments with limited visibility into concurrent obligations held by the same physical assets \cite{ForresterEtAl2023}.

This fragmentation can undermine enforcement in at least two ways. First, the same controllable capability may be credited toward multiple targets without de-rating for mutual exclusivity, creating risk that total credited capacity exceeds deliverable output when programs call simultaneously \cite{ShenotEtAl2024_PVPlusDERs}. Second, resources may be legitimately unavailable for one obligation because dispatched under another, yet system operators lack auditable records to verify priority or resolve conflicts, and standardized post-event processes for attributing delivery shortfalls across overlapping obligations remain limited \cite{ForresterEtAl2023}. 

\paragraph{Shortfall attribution under distribution network curtailment}

While Gate 3 deliverability assessments establish prospective deliverability under typical network conditions, real-time distribution constraints during scarcity hours may prevent delivery for reasons independent of resource availability. Many distribution operators now offer non-firm or flexible connection arrangements permitting interconnection but requiring curtailment when local limits bind. Implementation models include Active Network Management in Great Britain, Einspeisemanagement in Germany, flexible export limits in Australia, and Non-Firm Access in Ireland \cite{ESB_NonFirmAccess,monterde2025non}.

When distribution-directed curtailment coincides with bulk-system scarcity, prevailing pay-for-performance structures can expose aggregators to penalties for measured shortfalls without clearly distinguishing resource unavailability from network-directed curtailment \cite{ESB_NonFirmAccess,UK_CM_Flex_Response_2025}. This creates dual problems: aggregators face penalties for outcomes outside their control, suppressing enrollment in high-curtailment jurisdictions, while system operators may receive performance signals that confound resource reliability with network constraints. 
Both problems stem from the absence of auditable curtailment records and clear liability allocation between aggregators and network operators.

\subsubsection{Design directions}

\paragraph{Ex-ante obligation registries with priority hierarchies}

Rather than prohibiting multi-use or allowing unrestricted stacking, a unified obligation registry would track all capacity commitments for each DER portfolio across bulk-system RA, distribution services, ancillary markets, and retail programs \cite{FERC_Order_2222_2020,ForresterEtAl2023}. Each commitment is tagged with temporal scope, priority level, and mutual-exclusivity constraints, enabling operators to identify potential over-commitment and aggregators to price participation risk appropriately.

Implementation can be phased: (i) initial reporting maps stacking patterns; (ii) mandatory disclosure of primary reliability obligations during risk hours; and (iii) binding priority hierarchies determining which commitment prevails when calls overlap, such as bulk reliability over energy arbitrage or distribution safety over wholesale dispatch \cite{SEM_CRM_Performance_2022,SEM_16_051a_DeratingMethodology}. Order 2222 requires RTO-distribution coordination, while Ireland's reliability options and Great Britain's Capacity Market specify obligation hierarchies and force majeure provisions, though none fully integrate cross-program tracking with binding priority rules \cite{FERC_Order_2222_2020,Ofgem_CM_AnnualReport_2024_25,SEM_CRM_Performance_2022}. 

\paragraph{Real-time curtailment coordination with conditional excuse protocols}
Curtailment-aware settlement would create a basis for real-time coordination between distribution operations and bulk-system performance assessment  \cite{ESB_NonFirmAccess,UK_CM_Flex_Response_2025}. Such an approach would likely require three elements: (i) DSOs log curtailment instructions in time-synchronized auditable format recording affected assets, magnitude, and triggering conditions; (ii) settlement rules conditionally excuse verified curtailment up to specified thresholds while preserving aggregator incentives to maintain redundancy; and (iii) participation agreements clarify liability allocation, with aggregators responsible for understanding feeder-level limits and DSOs committed to transparent protocols \cite{ESIG_HighDER_Initiative_2022}.


\paragraph{Ex-post reconciliation with aggregation-level verification} 

Post-event settlement for multi-program participation would require reconciliation to attribute performance and resolve conflicts. While ex-ante registries define commitment hierarchies, actual dispatch during scarcity may deviate, requiring settlement without device-level traceability. An aggregation-level reconciliation ledger ties each scarcity hour to the binding obligation, metered portfolio-level quantity, and curtailment or program-call flags that adjust enforcement \cite{ACER_FrameworkGuideline_DR_2022}. 
Reconciliation enables two enforcement improvements: distinguishing resource unavailability from competing commitments, and verifying distribution-directed curtailment versus portfolio non-delivery during scarcity-hour shortfalls.

Supplementary Table~S1 records the jurisdictional examples discussed above and provides a compact gate-by-gate summary for readers.

\section{Cross-Stage Coupling and Coordination Requirements}
The gate-by-gate review identifies where DER capacity value is lost, but by itself it does not fully explain why targeted reforms have often failed to produce scalable outcomes. The central synthesis question is therefore not only where barriers arise, but how constraints established at one gate condition the choices available at the next. Addressing that question requires moving from gate-specific barriers to the linkages among institutional gates: rigidities at one gate propagate as binding constraints at others, such that solutions designed for individual gates remain incomplete without coordinated reform across the pathway. 
This section identifies three such linkages and derives the coordination principles they imply \cite{Brown2023_ESI_Requirements,leibowicz2024importance,ForresterEtAl2023}.

\subsection{Cross-Stage Coupling Mechanisms}
\begin{figure}[htbp]
\centering
\includegraphics[width=\textwidth]{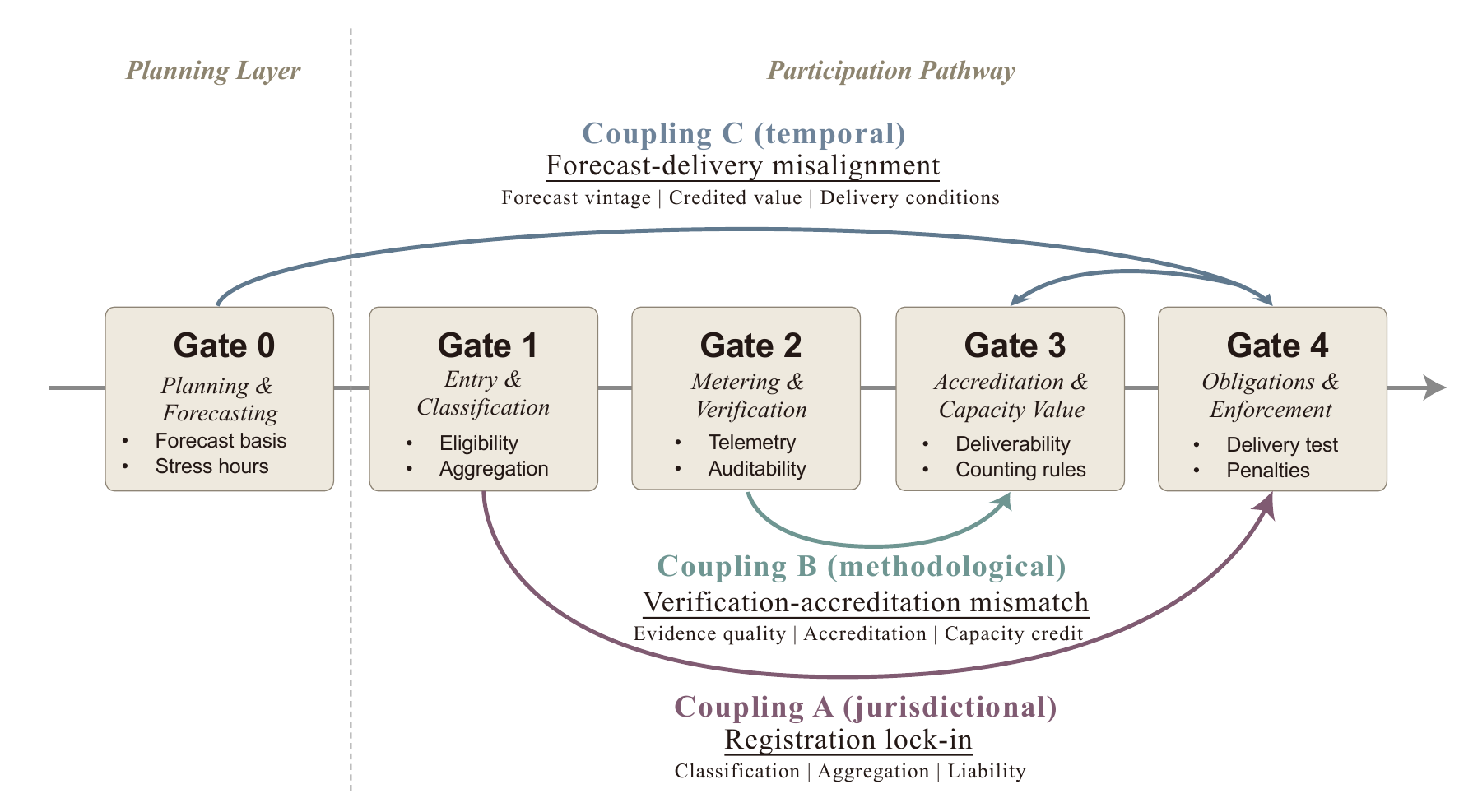}
\caption{Three cross-stage coupling mechanisms linking the stages of the DER adequacy pathway. Registration lock-in (G1$\leftrightarrow$G4), decoupling between verification and accreditation (G2$\leftrightarrow$G3), and misalignment between forecast vintages and accreditation or enforcement (G0$\leftrightarrow$G3/G4) each propagate institutional constraints across stages, explaining why isolated reforms often fail to produce scalable integration outcomes.}
\label{fig:crossgate_coupling}
\end{figure}

These linkages create three different coordination failures: a \emph{jurisdictional} boundary between entry rules and obligation structures (G1$\leftrightarrow$G4), a \emph{methodological} gap between verification methods and accreditation methodologies (G2$\leftrightarrow$G3), and a \emph{temporal} misalignment between planning assumptions and enforcement conditions (G0$\leftrightarrow$G3/G4). We focus on these three because each crosses a boundary between decision processes that single-gate reform cannot bridge on its own, rather than because they exhaust the linkages present in the pathway. Figure~\ref{fig:crossgate_coupling} illustrates how each coupling mechanism transmits institutional constraints across the pathway. 

\paragraph{Category registration propagates mismatched obligations (G1$\leftrightarrow$G4)}
The classification assigned at Gate~1 does not merely determine market access;  it strongly shapes the performance obligations enforced at Gate~4. A portfolio registered as demand response inherits event-based curtailment obligations designed for interruptible industrial loads, regardless of whether its actual composition includes state-dependent storage or weather-correlated generation \cite{Dykes2020_HybridPlants,DOE_2023_VPP_Liftoff}. This can create a self-reinforcing form of lock-in, whose persistence reflects a jurisdictional boundary rather than purely technical complexity. 
Classification rules operate under wholesale FERC/RTO authority, while retail contracts that shape actual obligations fall under state jurisdiction. 
Reforming G4 to recognize capability-based commitments requires G1 to first enable capability-based registration, yet wholesale-level G1 reform cannot compel corresponding changes to retail obligation structures across state jurisdictions. 
Order~2222 reveals this boundary clearly: it mandates RTO-level access for heterogeneous DER aggregations, yet explicitly preserves state authority to prohibit demand response from being bid into wholesale markets by third-party aggregators \cite{FERC_Order_2222_2020}. Aggregators must attest compliance with retail regulatory rules, but FERC cannot prescribe the content of those rules. 
European CRMs exhibit an analogous split: wholesale classification rules and distribution-level connection terms are governed by separate authorities, so that capacity market prequalification cannot constrain DSO curtailment decisions \cite{ESB_NonFirmAccess,UK_CM_Flex_Response_2025}. 
The result is expanded access within a categorical structure that remains mismatched to hybrid resource capabilities.

\paragraph{Verification granularity bounds accreditation (G2$\leftrightarrow$G3)}
The capacity value a resource can claim at Gate~3 is structurally bounded by the observability established at Gate~2, but the coupling mechanism is more specific than a simple data quality constraint. Class-level ELCC accreditation, the prevalent methodology in several major U.S. capacity markets, assigns identical capacity credits to all resources within a technology category regardless of individual performance \cite{pjm2021elcc,MISO_DLOL_2024}. This is a deliberate design choice: class-level averaging provides investment signal stability and administrative simplicity, but it also means that individual metering quality is largely irrelevant to accreditation outcomes by construction. A DER aggregation with high-resolution device-level telemetry receives the same capacity credit as one relying on aggregator-reported baselines, because the accreditation methodology pools both into the same class average. The structural trap is therefore not that better G2 data cannot be collected, but that G3 has no pathway to use it: transitioning to resource-level accreditation requires G2 to provide auditable individual-level performance records as a precondition, while investment in G2 upgrades remains economically irrational as long as G3 assigns class-level credits regardless of data quality. The same evidentiary limitation also propagates to enforcement: where Gate~2 cannot produce auditable individual-level records, Gate~4 settlement, performance attribution, and non-performance accountability inherit the same gap, so verification granularity ultimately bounds not only accreditation but enforceability \cite{Kua2023_AMI_PrivacySurvey,EU_Data_Act_2023}. In that sense, reforms at either gate are likely to remain limited unless the other changes as well.

\paragraph{Forecast vintage misaligns accreditation with enforcement conditions (G0$\leftrightarrow$G3 /G4)}
Gate~3 accreditation studies are anchored to net load forecasts and stress-period definitions established at Gate~0, while Gate~4 enforcement occurs against realized system conditions that may have diverged materially. The temporal gap is structural: procurement cycles operate on multi-year horizons by design, while passive BTM solar deployment evolves on sub-annual timescales \cite{nerc_ltra_2024,cec2025iepr}. 
Specifically, ELCC-based accreditation assigns capacity credits based on historically defined tight hours; when BTM penetration shifts net-load peak timing, these hours no longer correspond to Gate~4 enforcement triggers \cite{BestPractices_IRP_2024}.
Resources accredited against one set of tight-hour conditions may face performance obligations under later net-load peaks that differ materially from the conditions used to derive their credits. 
Adaptive refresh mechanisms that propagate observed BTM deployment into stress-period redefinition are structurally absent from current procurement frameworks, because forecast updates and accreditation recalculations operate on separate administrative cycles with no binding synchronization requirement. 
The misalignment can therefore become self-reinforcing \cite{BestPractices_IRP_2024}.

\subsection{Coordination Principles for Scalable Integration}

The three coupling mechanisms point to design principles that operate across gates by specifying how information, evidence quality, and timing assumptions move from one stage to the next. 

\paragraph{Codify cross-stage information handoffs (G1$\leftrightarrow$G4; G0$\leftrightarrow$G3/G4)}
Information flows between institutions are often informal or incomplete. Resolving the G1$\leftrightarrow$G4 boundary requires retail regulators to provide standardized notice of programs excluded from wholesale participation \cite{FERC_Order_2222_2020,ESIG_DER_Integration_2022}; resolving the G0$\leftrightarrow$G3/G4 timing gap requires forecast updates to enter accreditation models through codified procedures \cite{BestPractices_IRP_2024}. These handoffs can be embedded in tariffs and operating procedures through standardized outputs at defined intervals, converting informal expectations into enforceable obligations \cite{Brown2023_ESI_Requirements,ACER_FrameworkGuideline_DR_2022}.

\paragraph{Tie capacity credit to verification quality (G2$\leftrightarrow$G3)}
The G2$\leftrightarrow$G3 coupling cannot be resolved through observability improvements alone if accreditation ignores verification precision. MISO's FERC-approved DLOL methodology moves beyond pure class pooling through resource-level performance differentiation from the 2028/2029 planning year \cite{MISO_DLOL_2024}, but it still differentiates by historical availability rather than auditability. Breaking the coupling would require G3 methods to discount unauditable inputs and reward higher-precision verification, which in turn requires G2 to produce standardized precision declarations that accreditation models can consume \cite{dent2025resource,tarufelli2022capacity}. 

\paragraph{Integrate adaptive refresh mechanisms (G0$\leftrightarrow$G3/G4)}
California's Slice-of-Day framework disaggregates obligations across hours, but procurement targets remain anchored to forecasts finalized years earlier \cite{cpuc2025sod}. Adaptive refresh is therefore distinct from product disaggregation: it requires deviation thresholds for when observed BTM deployment triggers review, standardized data formats for moving forecast updates into accreditation models, and clear assignment of which institution initiates the refresh.

Partial antecedents exist, but they address different layers of the coordination problem: federal-state information exchange, RTO accreditation methodology, and planning-cycle administration. Scalable integration therefore depends on linking these layers through coordination obligations that current market designs do not yet require \cite{Brown2023_ESI_Requirements,ForresterEtAl2023,leibowicz2024importance}.

\section{Conclusions}
This review argues that the main barrier to translating distributed-resource capability into firm resource adequacy contributions is not DER technology alone, but the compliance architecture through which that capability is forecast, classified, verified, accredited, and enforced. Applying the five-stage pathway across California, PJM Interconnection, ISO New England, Great Britain, and Ireland reveals a recurring pattern: capacity value is often lost as distributed resources move from planning assumptions to participation categories, verification evidence, accredited capacity values, and performance obligations. The persistence of this pattern across different capacity mechanisms and regulatory structures suggests that many barriers to DER participation are structural features of RA design rather than jurisdiction-specific exceptions.

The framework's contribution is to make these losses visible as cross-stage coordination problems. Distributed resources do not enter RA frameworks only as physical assets with technical capabilities. They are translated, stage by stage, into institutional objects: forecast adjustments, eligible participation units, auditable performance records, capacity credits, and enforceable obligations. Each translation can narrow what counts, what can be measured, what can be credited, and what can be enforced. This helps explain why formal eligibility does not necessarily lead to scalable participation or reliable capacity credit.

The comparative synthesis identifies three coupling mechanisms through which constraints carry over across stages. Resource classification can determine later obligations and liability structures, making some hybrid or aggregated portfolios difficult to represent. Verification requirements can produce evidence that is too coarse, inconsistent, or difficult to audit for accreditation methods. Planning forecasts and credited values can also become misaligned with scarcity-hour performance as DER deployment and operating conditions change. These findings imply that RA reform should not focus only on relaxing individual participation requirements. More effective reform requires explicit information handoffs, auditable links between measured performance and accredited capacity, and mechanisms to refresh capacity values as deployment changes system conditions.

Future research should test whether reforms actually reduce cross-stage losses in capacity credit. Useful questions include whether higher-quality verification evidence changes accredited capacity values, whether new registration categories reduce constraints on hybrid portfolios, and whether capacity credits are updated when DER deployment changes system conditions. The central implication is therefore not that distributed resources should automatically receive more capacity credit. Rather, their contribution to adequacy depends on whether RA frameworks can consistently qualify, measure, credit, and enforce their capabilities. Distributed resources can become a dependable pillar of resource adequacy when these stages are designed to work together, so that physical capability is not lost at cross-stage interfaces.

\section*{Acknowledgements}
This work has been funded by the U.S. Department of Energy under the contract DE-AC02-05CH11231.

\section*{Declaration of competing interests}
The authors declare that they have no known competing financial interests or personal relationships that could have appeared to influence the work reported in this paper.

\section*{CRediT authorship contribution statement}
\textbf{Yujia Li:} Conceptualization, Methodology, Formal analysis, 
Writing: original draft.

\textbf{Alexandre Moreira:} Supervision, Writing: review \& editing.

\textbf{Miguel Heleno:} Supervision, Writing: review \& editing, Funding acquisition.

\section*{Data availability}
No new dataset was generated for this study. All materials analyzed in this review are cited in the reference list and supplementary material.

\bibliographystyle{elsarticle-num}
\bibliography{ra_refs_v3_clean} 

@techreport{doe2025resource,
  title     = {Resource Adequacy Report - Evaluating the Reliability and Security of the United States Electric Grid},
  author    = {{U.S. Department of Energy}},
  institution = {U.S. Department of Energy},
  year      = {2025},
  month     = {jul},
}

@techreport{nerc_ltra_2024,
  title        = {2024 Long-Term Reliability Assessment},
  author       = {{North American Electric Reliability Corporation}},
  institution  = {North American Electric Reliability Corporation},
  year         = {2024},
  month        = {December},
  url          = {https://www.nerc.com/pa/RAPA/ra/Reliability%20Assessments%20DL/NERC_Long%20Term%20Reliability%20Assessment_2024.pdf},
}

@techreport{doe2024future,
  title       = {The Future of Resource Adequacy},
  author      = {{U.S. Department of Energy}},
  institution = {U.S. Department of Energy},
  year        = {2024},
  month       = {apr},
}

@techreport{shehabi2024united,
  title={2024 United States Data Center Energy Usage Report},
  author={Shehabi, Arman and Newkirk, Alex and Smith, Sarah J and Hubbard, Alex and Lei, Nuoa and Siddik, Md Abu Bakar and Holecek, Billie and Koomey, Jonathan and Masanet, Eric and Sartor, Dale and others},
  institution={Lawrence Berkeley National Laboratory},
  year={2024},
  month={December},
}

@techreport{Forrester2021,
  title       = {Opportunities and Challenges to Capturing Distributed Battery Value via Retail Utility Rates and Programs},
  author      = {Forrester, Sydney P. and Cappers, Peter},
  institution = {Lawrence Berkeley National Laboratory},
  address     = {Berkeley, CA, USA},
  year        = {2021},
  DOI         = {10.2172/1835837},
  month       = dec,
}

@article{muhtadi2021distributed,
  title={Distributed energy resources based microgrid: Review of architecture, control, and reliability},
  author={Muhtadi, Abir and Pandit, Dilip and Nguyen, Nga and Mitra, Joydeep},
  journal={IEEE Transactions on Industry Applications},
  volume={57},
  number={3},
  pages={2223--2235},
  year={2021},
  DOI= {10.1109/TIA.2021.3065329},
  publisher={IEEE}
}

@misc{FERC_Order_2222_2020,
  author       = {{Federal Energy Regulatory Commission}},
  title        = {Order No.\ 2222: Participation of Distributed Energy Resource Aggregations in Markets Operated by Regional Transmission Organizations and Independent System Operators},
  year         = {2020},
  url          = {https://www.ferc.gov/sites/default/files/2020-09/E-1_0.pdf},
  note         = {Issued September 17, 2020}
}

@article{maheshwari2020effect,
  title={The effect of rate design on power distribution reliability considering adoption of distributed energy resources},
  author={Maheshwari, Aditya and Heleno, Miguel and Ludkovski, Michael},
  journal={Applied Energy},
  volume={268},
  pages={114964},
  year={2020},
  DOI={10.1016/j.apenergy.2020.114964},
  publisher={Elsevier}
}

@article{ochoa2023control,
  title={Control systems for low-inertia power grids: A survey on virtual power plants},
  author={Ochoa, Daniel E and Galarza-Jimenez, Felipe and Wilches-Bernal, Felipe and Schoenwald, David A and Poveda, Jorge I},
  journal={IEEE Access},
  volume={11},
  pages={20560--20581},
  year={2023},
  publisher={IEEE}
}

@ARTICLE{9701332,
  author={Zhou, Xiaoming and Sang, Maosheng and Bao, Minglei and Wang, Sheng and Cui, Wenqi and Ye, Chengjin and Ding, Yi},
  journal={IEEE Access}, 
  title={Exploiting Integrated Demand Response for Operating Reserve Provision Considering Rebound Effects}, 
  year={2022},
  volume={10},
  pages={15151-15162},
  doi={10.1109/ACCESS.2022.3148398}
}

@article{oikonomou2019deliverable,
  title={Deliverable energy flexibility scheduling for active distribution networks},
  author={Oikonomou, Konstantinos and Parvania, Masood and Khatami, Roohallah},
  journal={IEEE Transactions on Smart Grid},
  volume={11},
  number={1},
  pages={655--664},
  year={2019},
  doi={10.1109/TSG.2019.2927604},
  publisher={IEEE}
}

@book{billinton1970power,
  title={Power system reliability evaluation},
  author={Billinton, Roy},
  year={1970},
  publisher={Taylor \& Francis}
}

@article{garip2022power,
  title={Power System Reliability Assessment—A Review on Analysis and Evaluation Methods},
  author={Garip, Selahattin and {\"O}zdemir, {\c{S}}aban and Alt{\i}n, Necmi},
  journal={Journal of Energy Systems},
  volume={6},
  number={3},
  pages={401--419},
  year={2022},
  doi={10.30521/jes.1099618},
  publisher={Erol KURT}
}

@techreport{nerc2012methods,
  author       = {{NERC Integration of Variable Generation Task Force}},
  title        = {Methods to Model and Calculate Capacity Contributions of {Variable Generation} for {Resource Adequacy} Planning},
  institution  = {North American Electric Reliability Corporation},
  year         = {2011}
}

@techreport{nerc2012proba,
  author       = {{North American Electric Reliability Corporation}},
  title        = {2012 Probabilistic Assessment Report},
  institution  = {North American Electric Reliability Corporation},
  year         = {2012},
  month        = {November}
}

@techreport{stenclik2023ensuring,
  title={Ensuring Efficient Reliability: New Design Principles for Capacity Accreditation},
  author={Stenclik, Derek and Ahlstrom, Mark and Awara, Sarah and Bakke, Jordan and Bloom, Aaron and Cole, Wesley and Delaney, Elizabeth and Dobbie, Andrew and Elsevier, Sara and Figueroa-Acevedo, Armando and others},
  institution={Energy Systems Integration Group},
  year={2023},
  month={February},
}

@techreport{carvallo2023guide,
  title     = {A Guide for Improved Resource Adequacy Assessments in Evolving Power Systems: Institutional and technical dimensions},
  author    = {Carvallo, Juan Pablo and Zhang, Nan and Leibowicz, Benjamin D. and Carr, Thomas and Baik, Sunhee and Larsen, Peter H.},
  institution = {Lawrence Berkeley National Laboratory},
  year      = {2023},
  month     = {jun},
}

@article{botterud2020resource,
  title={Resource adequacy with increasing shares of wind and solar power: A comparison of european and us electricity market designs},
  author={Botterud, Audun and Auer, Hans},
  journal={Economics of Energy \& Environmental Policy},
  volume={9},
  number={2},
  pages={71--99},
  year={2020},
  publisher={International Association for Energy Economics}
}

@techreport{nerc_glossary_2016,
  author       = {{North American Electric Reliability Corporation}},
  title        = {Glossary of Terms Used in {NERC} Reliability Standards},
  institution  = {North American Electric Reliability Corporation},
  year         = {2025},
  month        = oct,
  note         = {Updated October 1, 2025},
}

@techreport{MISO2024,
  title       = {Resource Adequacy Metrics and Criteria Roadmap: A Reliability Imperative Report},
  author      = {{Midcontinent Independent System Operator (MISO)}},
  year        = {2024},
  month       = {12},
  address     = {Carmel, IN},
  type        = {Technical Report}
}

@article{wang2021crediting,
  title={Crediting variable renewable energy and energy storage in capacity markets: Effects of unit commitment and storage operation},
  author={Wang, Shen and Zheng, Ningkun and Bothwell, Cynthia D and Xu, Qingyu and Kasina, Saamrat and Hobbs, Benjamin F},
  journal={IEEE Transactions on Power Systems},
  volume={37},
  number={1},
  pages={617--628},
  year={2021},
  doi={10.1109/TPWRS.2021.3094408}, 
  publisher={IEEE}
}

@article{gonzato2023effect,
  title={The effect of short term storage operation on resource adequacy},
  author={Gonzato, Sebastian and Bruninx, Kenneth and Delarue, Erik},
  journal={Sustainable Energy, Grids and Networks},
  volume={34},
  pages={101005},
  year={2023},
  doi={10.1016/j.segan.2023.101005},
  publisher={Elsevier}
}

@techreport{caiso2025bpm,
  title={Business Practice Manual for Reliability Requirements},
  author={{California ISO}},
  institution={California ISO},
  year={2025},
  month={July},
}

@misc{CPUC_RA_Landing,
  author       = {{California Public Utilities Commission}},
  title        = {Resource Adequacy Compliance Materials},
  year         = {2025},
  howpublished = {\url{https://www.cpuc.ca.gov/industries-and-topics/electrical-energy/electric-power-procurement/resource-adequacy-homepage/resource-adequacy-compliance-materials}},
  note         = {Accessed March 14, 2026}
}

@techreport{nyiso2025icap,
  title={Installed Capacity Manual},
  author={{New York Independent System Operator (NYISO)}},
  institution={NYISO Capacity Market Products},
  year={2025},
  month={May},
}

@techreport{esig2021,
  author       = {{Redefining Resource Adequacy Task Force}},
  title        = {Redefining Resource Adequacy for Modern Power Systems: A Report of the Redefining Resource Adequacy Task Force},
  institution  = {Energy Systems Integration Group},
  year         = {2021},
  doi          = {10.2172/1961567}
}

@techreport{PJM_M18,
  title={PJM Manual 18: PJM Capacity Market},
  author={{PJM}},
  institution={PJM},
  year={2025},
  month={July},
  url={https://www.pjm.com/-/media/DotCom/documents/manuals/m18.ashx},
  note={Revision 62, effective December 17, 2025}
}

@techreport{caiso_lca_study_2026,
  author      = {{California ISO}},
  title       = {2026 Local Capacity Area Technical Study},
  year        = {2024},
  month       = {December},
  address     = {Folsom, CA}
}

@techreport{CAISO_TPD,
  author      = {{California Independent System Operator}},
  title       = {2024 Transmission Plan Deliverability Allocation Report},
  year        = {2024},
  month       = jun
}

@techreport{cpuc2025sod,
  author      = {{California Public Utilities Commission}},
  title       = {2025 Resource Adequacy and Slice of Day Guide},
  address     = {San Francisco, CA},
  year        = {2024},
  month       = sep
}

@techreport{pjm2021elcc,
  title        = {Manual 21A: Determination of Accredited UCAP Using Effective Load Carrying Capability Analysis},
  author       = {{PJM Interconnection}},
  organization = {PJM Interconnection},
  address      = {Audubon, PA},
  edition      = {Revision 5},
  month        = {June},
  year         = {2024},
  day          = {27}
}

@techreport{ISO_NE_M20_2023,
  author      = {{ISO New England}},
  title       = {{ISO} New England Manual for the Forward Capacity Market ({FCM})},
  number      = {Manual M-20, Revision 27},
  year        = {2023},
  month       = apr,
  url         = {https://www.iso-ne.com/static-assets/documents/2023/04/manual_20_forward_capacity_market_rev27_2023_04_06.pdf}
}

@misc{ISO_NE_Order2222_Conforming_2025,
  author       = {{ISO New England Inc.}},
  title        = {Revisions to Revise Rules for Distributed Energy Resource Aggregations, Docket No. ER26-105-000},
  year         = {2025},
  month        = oct,
  url          = {https://www.iso-ne.com/static-assets/documents/100028/order_2222_conforming_filing.pdf},
  note         = {Conforming filing for FERC Order No. 2222}
}

@misc{EU_Directive_2019_944,
  author       = {{European Union}},
  title        = {Directive (EU) 2019/944 on Common Rules for the Internal Market for Electricity and Amending Directive 2012/27/EU (Recast)},
  year         = {2019},
  note         = {OJ L 158, 14.6.2019, pp. 125--199}
}

@techreport{BestPractices_IRP_2024,
  author      = {Biewald, Bruce and Glick, Devi and Kwok, Shelley and Takahashi, Kenji
                 and Carvallo, Juan Pablo and Schwartz, Lisa C.},
  title       = {Best Practices in Integrated Resource Planning:
                 A Guide for Planners Developing the Electricity Resource Mix of the Future},
  institution = {Lawrence Berkeley National Laboratory and Synapse Energy Economics},
  year        = {2024},
  month       = nov,
  url         = {https://eta-publications.lbl.gov/sites/default/files/2024-12/irp_best_practices_2024_synapse_lbnl_24-061_0.pdf},
}

@techreport{cec2025iepr,
  title={2025 Integrated Energy Policy Report},
  author={{California Energy Commission}},
  institution={California Energy Commission},
  year={2025},
}

@techreport{cpuc_irp_2024,
  author      = {{California Public Utilities Commission}},
  title       = {Decision Adopting 2023 Preferred System Plan and Related Matters, and Addressing Two Petitions for Modification},
  number      = {Decision D.24-02-047},
  year        = {2024},
  month       = feb
}

@techreport{entsoe_mat_2023,
  author      = {{ENTSO-E}},
  title       = {European Resource Adequacy Assessment 2023},
  year        = {2023}
}

@misc{FERC_Order_2222A_2021,
  author       = {{Federal Energy Regulatory Commission}},
  title        = {Participation of Distributed Energy Resource Aggregations in Markets Operated by Regional Transmission Organizations and Independent System Operators, Order No. 2222-A},
  year         = {2021},
  note         = {174 FERC \textparagraph\ 61,197}
}

@techreport{ESIG_DER_Integration_2022,
  author       = {{Energy Systems Integration Group}},
  title        = {DER Integration into Wholesale Markets and Operations},
  year         = {2022},
  month        = {Jan.},
}

@techreport{JRC_ExplicitDR_2022,
  title        = {Explicit Demand Response for small end-users and independent aggregators: Status, context, enablers and barriers},
  author       = {Saviuc, I. and Zabala L{\'o}pez, C. and Pusk{\'a}s-Tompos, A. and Rollert, K. and Bertoldi, P.},
  institution  = {European Commission, Joint Research Centre},
  year         = {2022},
  number       = {EUR 31190 EN},
}

@techreport{CPUC_LIP_2024,
  author      = {{California Public Utilities Commission, Energy Division}},
  title       = {Load Impact Protocol Process Guide, Version 5.1},
  institution = {California Public Utilities Commission},
  year        = {2024},
  month       = dec
}

@misc{UK_CM_Flex_Response_2025,
  author       = {{Department for Energy Security and Net Zero}},
  title        = {Capacity Market: Proposals to Modernise Rules and Improve Participation and Delivery Assurance of Consumer-Led Flexibility -- Government Response},
  year         = {2025},
  note         = {Consultation outcome}
}

@techreport{Brown2023_ESI_Requirements,
  author       = {Brown, Richard E. and Khandekar, Aditya and Liu, Jingjing and Nordman, Bruce and Kolln, Jaime and Widergren, Steve and Narang, David and Bohn, Ted and Xue, Yaosuo},
  title        = {Energy Services Interface: Requirements Document},
  institution  = {Lawrence Berkeley National Laboratory},
  year         = {2023},
  month        = sep,
  doi          = {10.20357/B78S3H}
}

@techreport{EPRI_DERA_OrganizedMarkets_2021,
  author      = {Heidarifar, M. and Singhal, N. and Ela, E. and Lannoye, E. and Kristov, Lorenzo},
  title       = {Distributed Energy Resource Aggregation Participation in Organized Markets: {Federal Energy Regulatory Commission} Order 2222 Summary, Current State-of-the-Art, and Further Research Needs},
  institution = {Electric Power Research Institute (EPRI)},
  year        = {2021},
  address     = {Palo Alto, CA},
  number      = {3002020586}
}

@article{Huang2025_VPP_ResponseCapability,
  title={Review of virtual power plant response capability assessment and optimization dispatch},
  author={Huang, Junhui and Li, Hui and Zhang, Zhaoyun},
  journal={Technologies},
  volume={13},
  number={6},
  pages={216},
  year={2025},
  doi= {10.3390/technologies13060216},
  publisher={MDPI}
}

@article{Zhu2025_HybridEMS,
  title={A Review on Energy Management System for Grid-Connected Utility-Scale Renewable Hybrid Power Plants},
  author={Zhu, Rujie and Das, Kaushik and S{\o}rensen, Poul Ejnar and Hansen, Anca Daniela},
  journal={Wiley Interdisciplinary Reviews: Energy and Environment},
  volume={14},
  number={1},
  pages={e70004},
  year={2025},
  doi          = {10.1002/wene.70004},
  publisher={Wiley Online Library}
}

@techreport{Dykes2020_HybridPlants,
  title={Opportunities for research and development of hybrid power plants},
  author={Dykes, Katherine and King, Jennifer and DiOrio, Nicholas and King, Ryan and Gevorgian, Vahan and Corbus, David and Blair, Nate and Anderson, Kate and Stark, Greg and Turchi, Craig and others},
  year={2020},
  institution={National Renewable Energy Laboratory (NREL)},
  number       = {NREL/TP-5000-75026}
}

@article{DOE_2023_VPP_Liftoff,
  title={Pathways to commercial liftoff: Virtual power plants},
  author={Downing, Jennifer and Johnson, Nicholas and McNicholas, Mailinh and Nemtzow, David and Oueid, Rima and Paladino, Joseph and Wolfe, Elizabeth Bellis},
  journal={US Department of Energy report},
  year={2023}
}

@techreport{Horowitz2019_DERInterconnection,
  author       = {Horowitz, Kelsey A and Peterson, Zachary and Coddington, Michael H and Ding, Fei and Sigrin, Benjamin O and Saleem, Danish and Baldwin, Sara E and Lydic, Brian and Stanfield, Sky C and Enbar, Nadav and others},
  title        = {An Overview of Distributed Energy Resource ({DER}) Interconnection: Current Practices and Emerging Solutions},
  institution  = {National Renewable Energy Laboratory},
  year         = {2019},
  number       = {NREL/TP-6A20-72102},
  doi          = {10.2172/1508510}
}

@techreport{IEA2022_UnlockingDER,
  author       = {{International Energy Agency}},
  title        = {Unlocking the Potential of Distributed Energy Resources: Power System Opportunities and Best Practices},
  institution  = {International Energy Agency},
  year         = {2022},
  doi          = {10.1787/a2ed7a25-en}
}

@techreport{caiso2020bpmdr,
  author      = {{California Independent System Operator}},
  title       = {Business Practice Manual for Demand Response},
  institution = {California Independent System Operator},
  year        = {2026},
  note        = {Version 12},
  url         = {https://bpmcm.caiso.com/BPM%20Document%20Library/Demand%20Response/BPM_for_Demand_Response_V12_Clean.docx}
}

@techreport{CPUC_DER_Action_Plan_2022,
  title       = {Distributed Energy Resources Action Plan 2.0},
  author      = {{California Public Utilities Commission}},
  institution = {California Public Utilities Commission (CPUC)},
  year        = {2022},
  month       = apr,
}

@misc{EU_Implementing_Reg_2023_1162,
  author       = {{European Commission}},
  title        = {Commission Implementing Regulation (EU) 2023/1162 of 6 June 2023 on Interoperability Requirements and Non-Discriminatory and Transparent Procedures for Access to Metering and Consumption Data},
  year         = {2023},
  note         = {Official Journal of the European Union}
}

@article{Kua2023_AMI_PrivacySurvey,
  author       = {Kua, Jonathan and Hossain, Mohammad Belayet and Natgunanathan, Iynkaran and Xiang, Yong},
  title        = {Privacy Preservation in Smart Meters: Current Status, Challenges and Future Directions},
  journal      = {Sensors},
  year         = {2023},
  volume       = {23},
  number       = {7},
  pages        = {3697},
  doi          = {10.3390/s23073697}
}

@article{Psara2022_EU_DDS,
  author       = {Psara, Kyriaki and Papadimitriou, Christina and Efstratiadi, Marily and Tsakanikas, Sotiris and Papadopoulos, Panos and Tobin, Paul},
  title        = {European Energy Regulatory, Socioeconomic, and Organizational Aspects: An Analysis of Barriers Related to Data-Driven Services across Electricity Sectors},
  journal      = {Energies},
  year         = {2022},
  volume       = {15},
  number       = {6},
  pages        = {2197},
  doi          = {10.3390/en15062197}
}

@article{JorgensenMa2025_DT_EU,
  author       = {J{\o}rgensen, Bo N{\o}rregaard and Ma, Zheng Grace},
  title        = {Digital Twin of the European Electricity Grid: A Review of Regulatory Barriers, Technological Challenges, and Economic Opportunities},
  journal      = {Applied Sciences},
  year         = {2025},
  volume       = {15},
  number       = {12},
  pages        = {6475},
  doi          = {10.3390/app15126475}
}

@misc{EU_Data_Act_2023,
  author       = {{European Parliament} and {Council of the European Union}},
  title        = {Regulation ({EU}) 2023/2854 on Harmonised Rules on 
                  Fair Access to and Use of Data ({Data Act})},
  year         = {2023},
  month        = dec,
  howpublished = {Official Journal of the European Union, 
                  OJ L, 2023/2854, 22.12.2023},
}

@article{Papalexopoulos2024_VPP,
  title={Integrating Behind-the-Meter Grid Edge Technologies Into Wholesale Electricity Markets: A Novel Methodology Using Virtual Power Plants},
  author={Papalexopoulos, Alex and Oren, Shmuel and Chao, Hung-po},
  journal={IEEE Power and Energy Magazine},
  volume={23},
  number={1},
  pages={91--100},
  year={2025},
  month={Jan},
  publisher={IEEE},
  doi={10.1109/MPE.2024.3473852}
}

@article{Nikzad2024_AggregateDER,
  title={Estimating Aggregate Capacity of Connected {DER}s and Forecasting Feeder Power Flow With Limited Data Availability},
  author={Nikzad, Amir Reza and Mohamed, Amr Adel and Venkatesh, Bala and Penaranda, John},
  journal={IEEE Open Access Journal of Power and Energy},
  volume={11},
  pages={266--279},
  year={2024},
  month={Jun},
  publisher={IEEE},
  doi={10.1109/OAJPE.2024.3413606}
}

@article{dent2025resource,
  title={Resource adequacy and capacity Procurement: Metrics and decision support analysis},
  author={Dent, Chris J and Sanchez, Nestor and Shevni, Aditi and Smith, Jim Q and Wilson, Amy L and Yu, Xuewen},
  journal={Proceedings of the Institution of Mechanical Engineers, Part A: Journal of Power and Energy},
  volume={239},
  number={1},
  pages={199--209},
  year={2025},
  publisher={SAGE Publications Sage UK: London, England},
  doi  = {10.1177/09576509241292100}
}

@techreport{tarufelli2022capacity,
  title={Capacity markets for transactive energy systems},
  author={Tarufelli, Brittany L and Eldridge, Brent C and Somani, Abhishek},
  year={2022},
  institution={Pacific Northwest National Laboratory (PNNL), Richland, WA (United States)}
}

@techreport{Ofgem_CM_AnnualReport_2024_25,
  author       = {{Office of Gas and Electricity Markets (Ofgem)}},
  title        = {Annual Report on the Operation of the Capacity Market in 2024/25},
  institution  = {Ofgem},
  year         = {2025},
  month        = sep
}

@techreport{SEM_16_051a_DeratingMethodology,
  author       = {{EirGrid} and {SONI}},
  title        = {Appendix 1: TSOs Capacity Requirement and De-rating Factors Methodology},
  institution  = {SEM Committee},
  number       = {SEM-16-082a},
  year         = {2016},
  month        = dec
}

@techreport{PJM_M14B_2024,
  author      = {{PJM Interconnection}},
  title       = {{PJM} Manual 14B: {PJM} Region Transmission Planning Process},
  year        = {2025},
  month       = dec,
  note        = {Revision 58},
}

@techreport{CAISO_Deliv_Issue,
  author      = {{California Independent System Operator}},
  title       = {Generation Deliverability Methodology Review: 
                 Final Proposal},
  institution = {California Independent System Operator},
  year        = {2024},
  month       = jan
}

@techreport{ISONE_OP14_2025,
  author       = {{ISO New England Inc.}},
  title        = {Operating Procedure No. 14: Technical Requirements for Generators, Demand Response Resources, Asset Related Demands and Alternative Technology Regulation Resources},
  institution  = {ISO New England Inc.},
  year         = {2025},
  month        = apr,
  day          = {24},
  note         = {Revision 35}
}

@techreport{MISO_BPM011_2025,
  author      = {{Midcontinent Independent System Operator, Inc.}},
  title       = {Business Practices Manual {BPM-011}: Resource Adequacy},
  year        = {2024},
  month       = oct,
  note        = {Revision 30}
}

@techreport{CPUC_IRP_Incremental_ELCC_2023,
  author       = {{Energy and Environmental Economics, Inc.} and {Astrap{\'e} Consulting}},
  title        = {Incremental ELCC Study for Mid-Term Reliability Procurement: January 2023 Update},
  institution  = {Prepared for the California Public Utilities Commission},
  year         = {2023},
  month        = jan
}

@techreport{NYISO_DER_Aggregation_FAQ,
  author       = {{New York Independent System Operator}},
  title        = {Manual 04: Installed Capacity Manual},
  institution  = {New York Independent System Operator},
  year         = {2024},
  month        = may,
  day          = {31},
  note         = {Version 11.0}
}

@article{taheri2020fast,
  title={Fast probabilistic hosting capacity analysis for active distribution systems},
  author={Taheri, Sina and Jalali, Mana and Kekatos, Vassilis and Tong, Lang},
  journal={IEEE Transactions on Smart Grid},
  volume={12},
  number={3},
  pages={2000--2012},
  year={2020},
  doi          = {10.1109/TSG.2020.3038651},
  publisher={IEEE}
}

@article{cuenca2021non,
  title={Non-bias allocation of export capacity for distribution network planning with high distributed energy resource integration},
  author={Cuenca, Juan J and Hayes, Barry P},
  journal={IEEE Transactions on Power Systems},
  volume={37},
  number={4},
  pages={3026--3035},
  year={2021},
  publisher={IEEE},
  doi          = {10.1109/TPWRS.2021.3124999}
}

@techreport{Frick2021_LocationalValue,
  author      = {Frick, Natalie Mims and Price, Snuller and 
                 Schwartz, Lisa C. and Hanus, Nichole L. and 
                 Shapiro, Ben},
  title       = {Locational Value of Distributed Energy Resources},
  institution = {Lawrence Berkeley National Laboratory},
  address     = {Berkeley, CA},
  year        = {2021},
  month       = feb,
  number      = {LBNL-2001421}
}

@article{todd2019spillover,
  title={Spillover as a cause of bias in baseline evaluation methods for demand response programs},
  author={Todd, Annika and Cappers, Peter and Spurlock, C Anna and Jin, Ling},
  journal={Applied Energy},
  volume={250},
  pages={344--357},
  year={2019},
  publisher={Elsevier},
  doi          = {10.1016/j.apenergy.2019.05.050}
}

@techreport{smith2025potential,
  title={Potential Impacts of Dynamic Electricity Pricing in California: Load Shape and Customer Bill Impacts Under Elastic Customer Response},
  author={Smith, Sarah and Stuebs, Marius and Murthy, Samanvitha and Baik, Sunhee and Cappers, Peter and Gerke, Brian F and Brown, Richard E and Piette, Mary Ann},
  year={2025},
  institution={Lawrence Berkeley National Laboratory (LBNL), Berkeley, CA (United States)},
  doi          = {10.2172/2574291}
}

@article{Sun2023_CoolingPatterns_Snapback,
  title={Cooling-related electricity consumption patterns for small and medium businesses in California: Current impacts and future projections under climate change},
  author={Sun, Tao and Zanocco, Chad and Flora, June and Johnson, Samuel and Soto, Herie J and Rajagopal, Ram},
  journal={Energy and Buildings},
  volume={295},
  pages={113301},
  year={2023},
  publisher={Elsevier}
}

@techreport{NREL_DR_Fatigue_Snapback_2024,
  author      = {Xiong, Jie and Kim, Janghyun},
  title       = {End-Use Savings Shapes Measure Documentation: Thermostat Control for Load Shedding in Large Offices},
  institution = {National Renewable Energy Laboratory},
  address     = {Golden, CO},
  year        = {2024},
  month       = may,
  number      = {NREL/TP-5500-89340},
  type        = {Technical Report},
  doi          = {10.2172/2349284}
}

@techreport{CEC2019_SmartChargingEV,
  author       = {Black, Douglas and Yin, Rongxin and Wang, Bin},
  title        = {Smart Charging of Electric Vehicles and Driver Engagement for Demand Management and Participation in Electricity Markets},
  institution  = {California Energy Commission},
  address      = {Sacramento, CA},
  year         = {2019},
  number       = {CEC-500-2019-036},
}

@techreport{nethercutt2023resource,
  title={Resource Adequacy for State Utility Regulators: Current Practices and Emerging Reforms},
  author={Nethercutt, Elliott},
  year={2023},
  institution={National Association of Regulatory Utility Commissioners, Washington, DC}
}

@article{qi2023portfolio,
  title={Portfolio optimization of generic energy storage-based virtual power plant under decision-dependent uncertainties},
  author={Qi, Ning and Cheng, Lin and Li, Hongtao and Zhao, Yongliang and Tian, Hao},
  journal={Journal of energy storage},
  volume={63},
  pages={107000},
  year={2023},
  publisher={Elsevier},
  doi          = {10.1016/j.est.2023.107000}
}

@techreport{sun2021research,
  author      = {Sun, Yinong and Frew, Bethany and Levin, Todd and 
                 Hytowitz, Robin B. and Kwon, Jonghwan and 
                 Mills, Andrew D. and Xu, Qingyu and 
                 Heidarifar, Majid and Singhal, Nikita and 
                 de Mello, Phillip and Ela, Erik and 
                 Botterud, Audun and Zhou, Zhi and 
                 Hobbs, Ben F. and Crespo Montanes, Cristina},
  title       = {Research Priorities and Opportunities in {United States} Competitive Wholesale Electricity Markets},
  institution = {National Renewable Energy Laboratory},
  address     = {Golden, CO},
  year        = {2021},
  month       = may,
  number      = {NREL/TP-6A20-77521}
}

@article{carvallo2021implications,
  title={Implications of a regional resource adequacy program for utility integrated resource planning},
  author={Carvallo, Juan Pablo and Zhang, Nan and Leibowicz, Benjamin D and Carr, Thomas and Galbraith, Maury and Larsen, Peter H},
  journal={The Electricity Journal},
  volume={34},
  number={5},
  pages={106960},
  year={2021},
  publisher={Elsevier},
  doi          = {10.1016/j.tej.2021.106960}
}

@article{cole2020considerations,
  title={Considerations for maintaining resource adequacy of electricity systems with high penetrations of {PV} and storage},
  author={Cole, Wesley and Greer, Daniel and Ho, Jonathan and Margolis, Robert},
  journal={Applied Energy},
  volume={279},
  pages={115795},
  year={2020},
  publisher={Elsevier},
  doi={10.1016/j.apenergy.2020.115795}
}

@article{yuan2025flexibility,
  title={Flexibility Provision From Urban Buildings to Low-Carbon Power Systems: Quantification, Aggregation and System Integration},
  author={Yuan, Hening and Tang, Wenhu},
  journal={IET Energy Systems Integration},
  volume={7},
  number={1},
  pages={e70017},
  year={2025},
  publisher={Wiley Online Library}
}

@techreport{Keen2022_DistCapacityExpansion,
  title={Distribution capacity expansion planning: Current practice, opportunities, and decision support},
  author={Keen, Jeremy and Giraldez, Julieta and Cook, Elizabeth and Eiden, Andy and Placide, Scott and Hirayama, Alan and Monson, Brian and Mino, David and Eldali, Fathalla},
  year={2022}
}

@article{HCAReview2023,
  author  = {Umoh, Valentine and Davidson, Innes and Adebiyi, Adeola and Ekpe, Udo},
  title   = {Methods and Tools for PV and EV Hosting Capacity Determination in Low Voltage Distribution Networks --- A Review},
  journal = {Energies},
  year    = {2023},
  volume  = {16},
  number  = {8},
  pages   = {3609},
  doi     = {10.3390/en16083609},
}

@article{Bakare2023_DSM_AMI,
  title={A comprehensive overview on demand side energy management towards smart grids: challenges, solutions, and future direction},
  author={Bakare, Mutiu Shola and Abdulkarim, Abubakar and Zeeshan, Mohammad and Shuaibu, Aliyu Nuhu},
  journal={Energy Informatics},
  volume={6},
  number={1},
  pages={4},
  year={2023},
  publisher={Springer},
  doi          = {10.1186/s42162-023-00262-7}
}

@article{Sun2019_ResidentialBaseline,
  title={Clustering-based residential baseline estimation: A probabilistic perspective},
  author={Sun, Mingyang and Wang, Yi and Teng, Fei and Ye, Yujian and Strbac, Goran and Kang, Chongqing},
  journal={IEEE Transactions on Smart Grid},
  volume={10},
  number={6},
  pages={6014--6028},
  year={2019},
  publisher={IEEE}
}

@article{Lynch2019_DR_CapacityMarkets,
  author       = {Lynch, Muireann {\'A}. and Nolan, Sheila and Devine, Mel T. and O'Malley, Mark},
  title        = {The Impacts of Demand Response Participation in Capacity Markets},
  journal      = {Applied Energy},
  year         = {2019},
  volume       = {250},
  pages        = {444--451},
  doi          = {10.1016/j.apenergy.2019.05.063},
}

@misc{mantegna2024comprehensive,
  title        = {A Comprehensive Review of Capacity Accreditation Methods for Thermal, Variable Renewable, and Energy-limited Resources},
  author       = {Gianfranco Mantegna and Bethany Frew and Gregory Brinkman and Caitlin Murphy and Aaron Bloom and Wesley Cole},
  year         = {2024},
  eprint       = {2412.00533},
  archivePrefix = {arXiv},
  primaryClass = {eess.SY}
}

@article{leibowicz2024importance,
  title={The importance of capturing power system operational details in resource adequacy assessments},
  author={Leibowicz, Benjamin D and Zhang, Nan and Carvallo, Juan Pablo and Larsen, Peter H and Carr, Thomas and Baik, Sunhee},
  journal={Electric Power Systems Research},
  volume={228},
  pages={110057},
  year={2024},
  publisher={Elsevier},
  doi          = {10.1016/j.epsr.2023.110057}
}

@techreport{Ericson2022_HybridELCC,
  author       = {Olson, Arne and Ming, Zachary and Carron, Ben},
  title        = {{ELCC} Concepts and Considerations for Implementation},
  institution  = {Energy and Environmental Economics, Inc.},
  year         = {2021}
}

@techreport{ISO_NE_PFP_2025,
  author       = {{ISO New England Inc.}},
  title        = {Performance of Capacity Resources and Pay for Performance},
  type         = {Memorandum to NEPOOL Markets Committee},
  year         = {2022},
  month        = sep,
}

@techreport{UK_CM_Scheme_Update_2025,
  author       = {{Department for Energy Security and Net Zero}},
  title        = {Contracts for Difference and Capacity Market Scheme Update 2025},
  institution  = {Department for Energy Security and Net Zero},
  address      = {London, UK},
  year         = {2025},
  month        = dec,
  number       = {HC 1590},
  isbn         = {978-1-5286-6143-0}
}

@techreport{SEM_CRM_Performance_2022,
  author       = {{Single Electricity Market Committee}},
  title        = {Performance of the SEM Capacity Remuneration Mechanism},
  institution  = {Single Electricity Market Committee},
  number       = {SEM-22-054A},
  year         = {2022},
  month        = jun
}

@techreport{ShenotEtAl2024_PVPlusDERs,
  author       = {Shenot, John and Linvill, Carl and Dupuy, Max and Brutkoski, Donna},
  title        = {Capturing More Value from Combinations of {PV} and Other Distributed Energy Resources},
  institution  = {National Renewable Energy Laboratory},
  type         = {Subcontract Report},
  number       = {NREL/SR-7A40-90129},
  year         = {2019},
  doi          = {10.2172/2394648}
}

@article{gulotta2023opening,
  title={Opening of ancillary service markets to distributed energy resources: A review},
  author={Gulotta, Francesco and Dacco, Edoardo and Bosisio, Alessandro and Falabretti, Davide},
  journal={Energies},
  volume={16},
  number={6},
  pages={2814},
  year={2023},
  publisher={MDPI},
  doi          = {10.3390/en16062814}
}

@techreport{ForresterEtAl2023,
  author       = {Forrester, Sydney P. and Triedman, Cole and Kozel, Sam and Brooks, Cameron and Cappers, Peter},
  title        = {Third-Party Aggregation Rulemaking in MISO and SPP Footprints},
  institution  = {Lawrence Berkeley National Laboratory},
  address      = {Berkeley, CA, USA},
  year         = {2023},
  month        = sep
}

@misc{ACER_FrameworkGuideline_DR_2022,
  author       = {{European Union Agency for the Cooperation of Energy Regulators (ACER)}},
  title        = {Framework Guideline on Demand Response},
  year         = {2022},
  note         = {Official framework guideline}
}

@techreport{ESB_NonFirmAccess,
  author       = {{ESB Networks}},
  title        = {Guide - Non-Firm Access Connections for Distribution Connected Distributed Generators},
  institution  = {ESB Networks},
  address      = {Dublin, Ireland},
  year         = {2021},
  month        = may
}

@techreport{ESIG_HighDER_Initiative_2022,
  author       = {{Energy Systems Integration Group}},
  title        = {The Transition to a High-DER Electricity System: Creating a National Initiative on DER Integration for the United States},
  institution  = {Energy Systems Integration Group},
  address      = {Reston, VA},
  year         = {2022},
  month        = aug
}

@article{monterde2025non,
  author       = {Monterde, Manuel Romeo and Alvarez, Erik F. and Valarezo, Orlando},
  title        = {Non-Firm Grid Connections: A Review of Access Types, Mechanisms, and Regulatory Frameworks},
  journal      = {Current Sustainable/Renewable Energy Reports},
  year         = {2025},
  volume       = {12},
  number       = {1},
  pages        = {23},
  doi          = {10.1007/s40518-025-00268-7}
}

@ARTICLE{5374085,
  author={Leite da Silva, Armando M. Leite and Sales, Warlley S. and Manso, Luiz AntÔnio da Fonseca and Billinton, Roy},
  journal={IEEE Transactions on Power Systems}, 
  title={Long-Term Probabilistic Evaluation of Operating Reserve Requirements With Renewable Sources}, 
  year={2010},
  volume={25},
  number={1},
  pages={106-116},
  doi={10.1109/TPWRS.2009.2036706}}

@techreport{MISO_DLOL_2024,
  author      = {{Midcontinent Independent System Operator (MISO)}},
  title       = {Resource Accreditation Reform: Direct Loss-of-Load Methodology},
  year        = {2024},
  month       = mar,
  note        = {FERC Docket No. ER24-1638; approved October 2024, effective 2028/2029 planning year},
}

@article{Gao2024_VPP_APEN,
  author       = {Gao, Hongchao and Jin, Tai and Feng, Cheng and Li, Chuyi and Chen, Qixin and Kang, Chongqing},
  title        = {Review of virtual power plant operations: Resource coordination and multidimensional interaction},
  journal      = {Applied Energy},
  volume       = {357},
  pages        = {122284},
  year         = {2024},
  doi          = {10.1016/j.apenergy.2023.122284},
  url          = {https://doi.org/10.1016/j.apenergy.2023.122284}
}

@article{Haugen2024_MarketModels_APEN,
  author       = {Haugen, Mari and Blaisdell-Pijuan, Paris L. and Botterud, Audun and Levin, Todd and Zhou, Zhi and Belsnes, Michael and Korp{\aa}s, Magnus and Somani, Abhishek},
  title        = {Power market models for the clean energy transition: State of the art and future research needs},
  journal      = {Applied Energy},
  volume       = {357},
  pages        = {122495},
  year         = {2024},
  doi          = {10.1016/j.apenergy.2023.122495},
  url          = {https://doi.org/10.1016/j.apenergy.2023.122495}
}

@article{Menci2024_LFM_APEN,
  author       = {Menci, Sergio Potenciano and Valarezo, Orlando},
  title        = {Decoding design characteristics of local flexibility markets for congestion management with a multi-layered taxonomy},
  journal      = {Applied Energy},
  volume       = {357},
  pages        = {122203},
  year         = {2024},
  doi          = {10.1016/j.apenergy.2023.122203},
  url          = {https://doi.org/10.1016/j.apenergy.2023.122203}
}

@misc{EirGrid_DSU_2026,
  author       = {{EirGrid}},
  title        = {DSU Setup and Testing},
  year         = {2026},
  url          = {https://www.eirgrid.ie/customer-and-industry/general-customer-information/grid-code-compliance-test/compliance-testing/dsu},
  note         = {Accessed 2026-04-10}
}

@techreport{ISONE_AMR_2024,
  author       = {{ISO New England}},
  title        = {2024 Annual Markets Report},
  institution  = {ISO New England Inc.},
  year         = {2025},
  url          = {https://www.iso-ne.com/static-assets/documents/100023/2024-annual-markets-report.pdf},
  note         = {Accessed 2026-04-10}
}

@article{Valentini2022_CBL_Review,
  author       = {Valentini, Ottavia and Andreadou, Nikoleta and Bertoldi, Paolo and Lucas, Alexandre and Saviuc, Iolanda and Kotsakis, Evangelos},
  title        = {Demand Response Impact Evaluation: A Review of Methods for Estimating the Customer Baseline Load},
  journal      = {Energies},
  year         = {2022},
  volume       = {15},
  number       = {14},
  pages        = {5259},
  doi          = {10.3390/en15145259}
}

@techreport{EldridgeSomani2022_FERC2222,
  author       = {Eldridge, Brent C. and Somani, Abhishek},
  title        = {Impact of {FERC} Order 2222 on {DER} Participation Rules in {US} Electricity Markets},
  institution  = {Pacific Northwest National Laboratory (PNNL)},
  year         = {2022},
  number       = {PNNL--33383},
  doi          = {10.2172/1968796}
}

@article{Tillmanns2026_ProbabilisticRA,
  author       = {Tillmanns, Maximilian and Sch{\"o}ttler, Julia and Praktiknjo, Aaron},
  title        = {A review of probabilistic resource adequacy assessments in power systems: Methods, applications, and future challenges},
  journal      = {Energy Policy},
  year         = {2026},
  volume       = {209},
  pages        = {114924},
  doi          = {10.1016/j.enpol.2025.114924}
}

@article{SanchezJimenez2025_CRMs_APEN,
  author       = {Sanchez Jimenez, Ingrid and Bruninx, Kenneth and de Vries, Laurens J.},
  title        = {Capacity remuneration mechanisms for decarbonized power systems},
  journal      = {Applied Energy},
  year         = {2025},
  volume       = {391},
  pages        = {125878},
  doi          = {10.1016/j.apenergy.2025.125878}
}

@article{Liu2024_VPP_DynamicAggregation,
  title   = {Dynamic aggregation strategy for a virtual power plant to improve flexible regulation ability},
  author  = {Liu, Xin and Li, Yang and Wang, Li and Tang, Junbo and Qiu, Haifeng and Berizzi, Alberto and Ilea, Valentin and Gao, Ciwei},
  journal = {Energy},
  volume  = {297},
  pages   = {131261},
  year    = {2024},
  publisher = {Elsevier}
}

@article{WenSong2023_WindStorage_ELCC,
  title   = {{ELCC}-based capacity value estimation of combined wind--storage system using {IPSO} algorithm},
  author  = {Wen, Lei and Song, Qianqian},
  journal = {Energy},
  volume  = {263},
  pages   = {125784},
  year    = {2023},
  publisher = {Elsevier},
  doi     = {10.1016/j.energy.2022.125784}
}

@article{Wang2024_MultiTimeScale_CapCredit,
  title   = {Multi-time-scale capacity credit assessment of renewable and energy storage considering complex operational time series},
  author  = {Wang, Renshun and Wang, Shilong and Geng, Guangchao and Jiang, Quanyuan},
  journal = {Applied Energy},
  volume  = {355},
  pages   = {122382},
  year    = {2024},
  publisher = {Elsevier},
  doi     = {10.1016/j.apenergy.2023.122382}
}

\includepdf[pages=-,landscape,fitpaper]{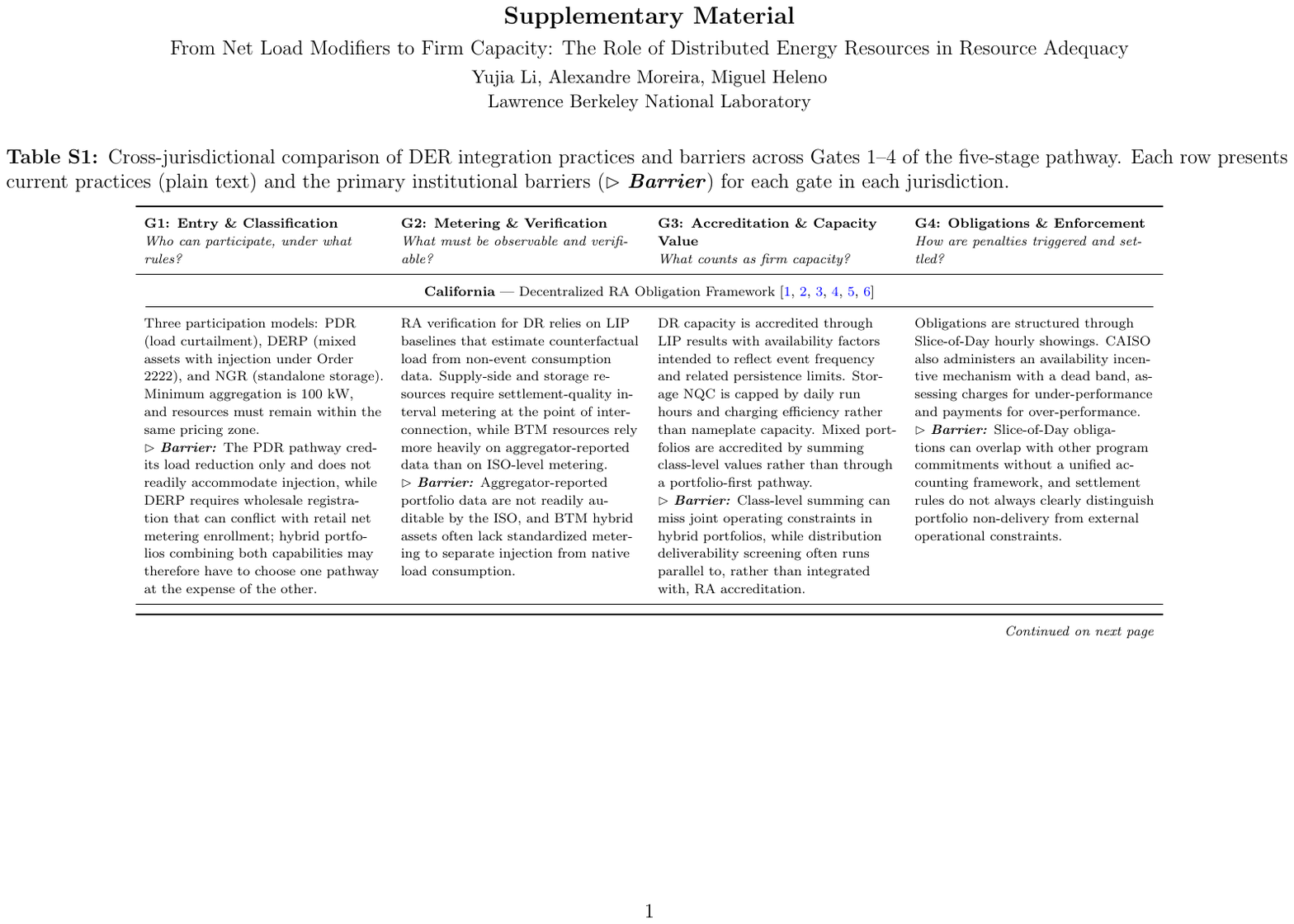}

\end{document}